\documentclass[prb,aps,tightenlines,twocolumn,superscriptaddress,showpacs,preprintnumbers,citeautoscript,11pt]{revtex4-1} 

\usepackage{amsmath}
\usepackage{amssymb}
\usepackage{mathrsfs}
\usepackage{bm}
\usepackage{graphicx}
\usepackage{xfrac}
\usepackage{url}
\usepackage{tikz}
\usepackage{multirow}
\usepackage{mathtools}
\usepackage{braket}
\usepackage{courier}
\usepackage[version=4]{mhchem}
\usepackage[T1]{fontenc}
\usepackage{scalerel}
\usepackage{soul}
\usepackage{algpseudocode}
\usepackage[ruled]{algorithm2e}
\usepackage{tabularx}
\usepackage{xspace}
\usepackage{hyperref}

\begin{document}

\title{First-Principles Atomistic Structure and Dynamics of Polyethylene During High-Pressure Radical Polymerization via Machine Learning Force Fields}

\author{Bharatha~K.~Gunawardana}
\altaffiliation{These authors contributed equally to this work.}
\affiliation{Department of Chemistry, University of North Texas, Denton, TX 76203, USA}
\author{Teresa~Shah}
\altaffiliation{These authors contributed equally to this work.}
\affiliation{Department of Chemistry, University of North Texas, Denton, TX 76203, USA}
\author{Bicha~Azizova}
\affiliation{Department of Chemical and Biomolecular Engineering, Lehigh University, Bethlehem, PA 18015, USA}
\author{Deepa~Ranabhat}
\affiliation{Department of Chemistry, University of North Texas, Denton, TX 76203, USA}
\author{Yizhi~Song}
\affiliation{Department of Chemistry, University of North Texas, Denton, TX 76203, USA}
\author{Akshath~Shastri}
\affiliation{Department of Chemistry, University of North Texas, Denton, TX 76203, USA}
\author{Srinjoy~Ghose}
\affiliation{Department of Chemistry, University of North Texas, Denton, TX 76203, USA}
\author{Thomas~E.~Gartner~III}
\email{teg323@lehigh.edu}
\affiliation{Department of Chemical and Biomolecular Engineering, Lehigh University, Bethlehem, PA 18015, USA}
\author{Hsin-Yu~Ko}
\email{hsin-yu.ko@unt.edu}
\affiliation{Department of Chemistry, University of North Texas, Denton, TX 76203, USA}

\date{\today}

\begin{abstract}
Polyethylene (PE) is one of the most commonly used synthetic polymers. While the synthesis and processing protocols for PE are well established, precise experimental assignment of microscopic structures at atomistic resolution (i.e., the position of each atom) remains largely limited to highly crystalline systems.
This gap is often addressed via computer simulations using empirical interatomic potentials, which use approximate but efficient descriptions of interatomic interactions to reach the length and time scales needed to describe macromolecules.
These empirical potentials typically perform well for bulk and/or collective properties but face challenges with chemical realism for complex systems, e.g., during reactive processes.
In this work, we address this challenge by combining the computational efficiency of a deep potential (DP) machine-learning force field and the chemical realism of first-principles van der Waals (vdW) corrected hybrid density functional theory (DFT) enabled by a \texttt{SeA} high-throughput framework.
Using this approach, we study the structure and dynamics of PE oligomers and polymers in an ethylene solvent under common high-pressure (supercritical) radical polymerization conditions.
We found that the local solvation environment of radical-containing PE oligomers converges for chain lengths greater than ($n\approx 6$), suggesting extensibility of our oligomer-trained MLFF to significantly longer polymers.
We then confirmed the extensibility of these models to long PE chains by characterizing the molecular weight scaling of single-chain structure and dynamics, which showed classic good solvent behavior.
Our PE MLFF retained a consistent level of fidelity and stability across a wide range of thermodynamic state points and chain lengths, at full atomistic resolution, therefore paving the way towards first-principles-based polymer structure and property prediction.
\end{abstract}

\maketitle

\section{Introduction}

Polymers are a unique class of materials that take center stage in many applications of chemistry, physics, and biology, owing to their light weight, toughness, and resistance to degradation.
While the synthesis-processing-property relationships of many polymers have been well-established, experimental assignment of the 3D atomistic structures of amorphous and semi-crystalline polymers at the angstrom-to-nanometer scale remains largely limited to angularly averaged radial distributions obtained from X-ray and neutron scattering due to their (in general) lack of long-range order.

As a result, computer simulations (e.g., molecular dynamics (MD)~\cite{allen_computer_1989,frenkel_understanding_2001} and/or multiscale simulations~\cite{zeng_multiscale_2008,engquist_multiscale_2009,van_der_giessen_roadmap_2020,gartner_modeling_2019}) have been instrumental in understanding the structure, dynamics, and collective properties of polymers.
However, the accuracy and utility of these simulations rely heavily on the quality of the underlying empirical force fields used to model the interatomic interactions~\cite{gartner_modeling_2019}.
Traditional empirical force fields have found success in modeling many classes of common polymers, but typically cannot capture phenomena such as reactivity, polarizability, strong polymer-ion interactions, or other effects sensitive to the underlying electronic structure.
For these classes of problems, which are increasingly prevalent at the forefront of polymer science and engineering, one approach to address this issue would be to perform \textit{ab initio} MD (AIMD) simulations, which compute the atomic forces from first-principles quantum electronic structure theory~\cite{car_unified_1985,marx_ab_2009}.
In doing so, the level of chemical detail included in the prediction of atomic forces can be improved systematically as one applies increasingly more advanced levels of electronic structure theory, widening the set of advanced phenomena that can be captured in polymer simulations.
Such an approach could provide access to direct simulations of topics as, e.g., polymer degradation and (de)polymerization~\cite{mieda_comparison_2025,ma_understanding_2023}, conjugated, charged, and/or radical-containing polymers~\cite{li_modeling-driven_2024,ma_perspective_2021,tan_bridging_2023}, and mixed ion-electron conducting polymers~\cite{qin_organic_2025}.
However, standard AIMD (or DFT)  simulations of polymers, even on the latest hardware, are simply not compatible with the length and time scales needed to capture macromolecular phenomena~\cite{gartner_modeling_2019}, therefore resulting in only a small number of polymer AIMD studies; most of these focus on highly crystalline systems for computational tractability~\cite{bernasconi_solid-state_1997,boero_first_2000,ferretti_ab_2004,beyer_mechanochemistry_2005,fontana_high-pressure_2007,ribas-arino_covalent_2012,liu_how_2012,olsson_ab_2017,kurita_crystalline_2018,xue_ab_2019,huan_polymer_2020}.

To achieve a reliable description of the quantum electronic interactions, hybrid~\cite{becke_density-functional_1993} density functional theory (DFT)~\cite{hohenberg_inhomogeneous_1964,kohn_self-consistent_1965,jones_density_1989,parr_density-functional_1989} has been demonstrated to provide semi-quantitative accuracy in previous gas-phase studies of oligomer species~\cite{salzner_design_1997,salzner_accurate_1998,salzner_comparison_1998,vaschetto_first-principles_1999,de_oliveira_energy_2000,zade_short_2011,korzdorfer_organic_2014}.
In condensed phases, given the importance of long-range van der Waals (vdW) interactions~\cite{grimme_dispersion-corrected_2016,hermann_first-principles_2017} in determining polymer properties, these effects also need to be corrected~\cite{hong_first-principles-based_2021}.
The use of (vdW-corrected) hybrid DFT requires the evaluation of the exact Hartree--Fock exchange (EXX) interaction, which is typically computationally prohibitive for large-scale condensed-phase systems when using the conventional convolution theorem based approach~\cite{gygi_self-consistent_1986} due to its cubic-scaling cost with system size.
However, the computational burden associated with hybrid DFT is substantially decreased by a recently developed \texttt{SeA} high-throughput framework by harnessing three-levels of savings originated from the natural sparsity of the exchange interaction in real space within a localized orbital representation~\cite{ko_high-throughput_2023}.
The \texttt{SeA} framework harnesses three levels of computational savings by seamlessly combining the selected columns of the density matrix (SCDM)~\cite{damle_compressed_2015} orbital localization scheme, a black-box linear-scaling EXX engine within a localized orbital representation (\texttt{exxl}, derived from Refs.~\onlinecite{ko_enabling_2020,ko_enabling_2021}), and an adaptively compressed exchange (ACE) operator~\cite{lin_adaptively_2016}.
In doing so, \texttt{SeA} performs hybrid DFT calculations at a cost comparable to within a small prefactor (around $4\times$$\mathrm{-}$$6\times$) of the commonly used generalized-gradient approximations (GGA)~\cite{becke_density-functional_1988,lee_development_1988,jones_density_1989,perdew_comparison_1996,perdew_generalized_1996,perdew_jacobs_2001}, which are less reliable than hybrid DFT but are commonly applied in condensed-phase simulations due to their reduced computational cost.
The accuracy of vdW-corrected hybrid DFT opens the door to simulating the reactive processes that underlie polymer synthesis, paving the way to directly model polymerization for non-empirical, bottom-up prediction of polymer structure and properties.

However, even with state-of-the-art electronic structure methods, the DFT calculations to drive AIMD simulations on systems of the requisite size to capture polymeric phenomena are still intractable. Thus, researchers are beginning to explore the development of machine-learning force fields (MLFFs)~\cite{deringer_machine_2019,behler_four_2021,ceriotti_introduction_2021,ko_general-purpose_2021,unke_machine_2021,meuwly_machine_2021,huang_ab_2021,kocer_neural_2022,wen_deep_2022} as an additional improvement in computational cost.
In this approach, one trains an MLFF to reproduce the atomic energies and forces as predicted by an electronic-structure method of choice, thus enabling first-principles simulations at larger length and time scales than the reference first-principles method.
While MLFFs have quickly become popular in materials simulations, their application to polymers remains comparatively less developed.
Recent studies have begun to explore their effectiveness for predicting polymer properties~\cite{hong_first-principles-based_2021}. These models have enabled first-principles prediction of polymer-relevant structural, transport, thermophysical, and mechanical observables, with reported validation against experimental or \textit{ab initio} benchmarks~\cite{hong_first-principles-based_2021,mohanty_development_2023,chen_phyneo_2023,mieda_comparison_2025,simm_simpoly_2025}. Applications now include both chemically-specific polymers and broader polymer families, indicating growing but still limited transferability across polymer chemistries~\cite{wang_scalable_2021,long_polymers_2024,simm_simpoly_2025,hooven_how_2026}. A key complication is that polymer properties depend on local monomer chemistry, chain conformation, topology, and interchain organization across multiple length and time scales. This multiscale coupling makes long-range and non-covalent interactions, representative polymer datasets, stable long-time dynamics, and model interpretability central unresolved issues for polymer MLFFs~\cite{gartner_modeling_2019,schmid_understanding_2023,hu_efficient_2024}.

In this work, we applied the \texttt{SeA} framework to address the challenges of quantum electronic structure prediction in polymers, and used the resulting DFT data to train a deep potential (DP)~\cite{zhang_deep_2018,zhang_end--end_2018} MLFF.
With sufficient training data, DP molecular dynamics (DPMD) provides $10^6\times$$\mathrm{-}$$10^8\times$ speedup at large length- and time-scales without loss of DFT accuracy,~\cite{zhang_deep_2018,zhang_end--end_2018,ko_isotope_2019} therefore significantly lowering the computational cost for simulating macromolecules. Critically, once trained, the computational cost of DPMD scales nearly linearly with system size, so the advantage of a DP compared to DFT only improves as one approaches macromolecular length scales. 
Here, we investigated the structure of PE oligomer radicals in an ethylene solvent as a function of chain length under high-pressure polymerization conditions ($T \geq 200$~$^\circ$C and $p \geq 500$~bar)~\cite{aggarwal_polyethylene_1957}.
Based on radial distribution functions (RDFs) and localized orbitals, we demonstrate the rapid convergence of both the nuclear and electronic solvation environments around the radical site with respect to chain length, revealing the possibility of leveraging the oligomer local environment to model the macromolecular limit.
Further, we show the extent to which an MLFF trained on oligomeric species can extrapolate to polymer solutions, key knowledge needed to design useful polymer MLFFs~\cite{hooven_how_2026}. We examine the molecular weight scaling of PE chain configurations and dynamics in dilute solution, and demonstrate that our oligomer-trained MLFF provides reasonable descriptions of large-scale polymer properties. These results underscore the capability of our modeling approach to characterize polymer behavior across the atomic to macromolecular length scales.

\section{Training the Deep Potential Machine-Learning Force Field}
\begin{figure}[ht!]
    \centering
    \includegraphics[width=\linewidth]{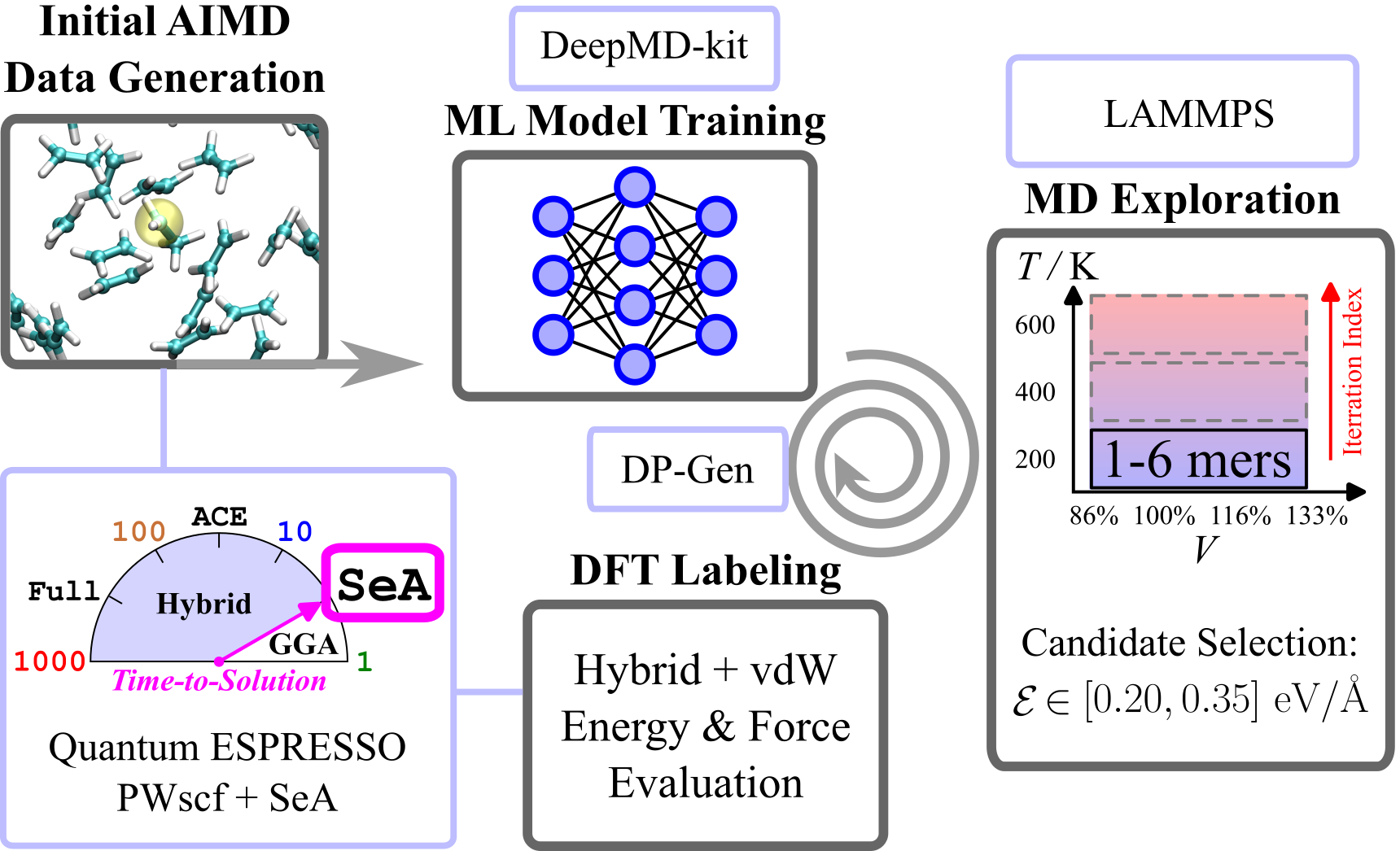}
    \caption{Schematic illustration of our workflow for training the machine-learning force field (MLFF) at the hybrid DFT+vdW level by integrating \texttt{SeA} and the DP ecosystem.
    }
    \label{fig:ml_scheme}
\end{figure}
To study the atomistic structures of PE during radical polymerization, we trained a DP MLFF~\cite{zhang_deep_2018,zhang_end--end_2018} with the PBE0+D3 functional~\cite{perdew_rationale_1996,adamo_toward_1999,grimme_consistent_2010} by integrating the \texttt{SeA} hybrid DFT engine with the DP software ecosystem (Fig.~\ref{fig:ml_scheme}). This combination of approaches allows us to rapidly generate MLFF training data using the comparatively large system sizes necessary for polymers, while maintaining the chemical accuracy of a hybrid DFT level of electronic structure theory.
We assembled the MLFF training dataset by starting from short AIMD simulations of oligomer radicals with chain lengths ($n$, number of \ce{C2H4} structural units) ranging from $n=1$ (i.e., \ce{C2H5*}) to $n=6$ (i.e., \ce{C12H27*}), covering combinations of temperatures of $T\in\{200,400,600\}$~K and densities of $\rho\in\{0.43, 0.49,  0.57, 0.66\}$~g/cm$^3$. Then, we performed \texttt{DP-Gen} active learning cycles~\cite{zhang_active_2019,zhang_dp-gen_2020} involving iterative MD explorations of the same set of oligomers and thermodynamic range (see Appendix~\ref{app:compt_details} for detailed methods).
During the active machine learning process, we intentionally introduced small unit cells and leveraged the presence of nearby periodic images of the oligomer chains to encode chain--chain interactions in our training dataset. The completed active learning process led to $99,074$ labeled PBE0+D3 structures, which was sufficient to produce a stable and accurate DP MLFF, the properties of which we explore in the following sections.  

\section{Thermodynamic Properties and Solvation Structure of The Ethyl Radical \label{sec:ethyl_rad}}

\begin{figure*}[ht!]
    \centering
    \includegraphics[width=0.8\linewidth]{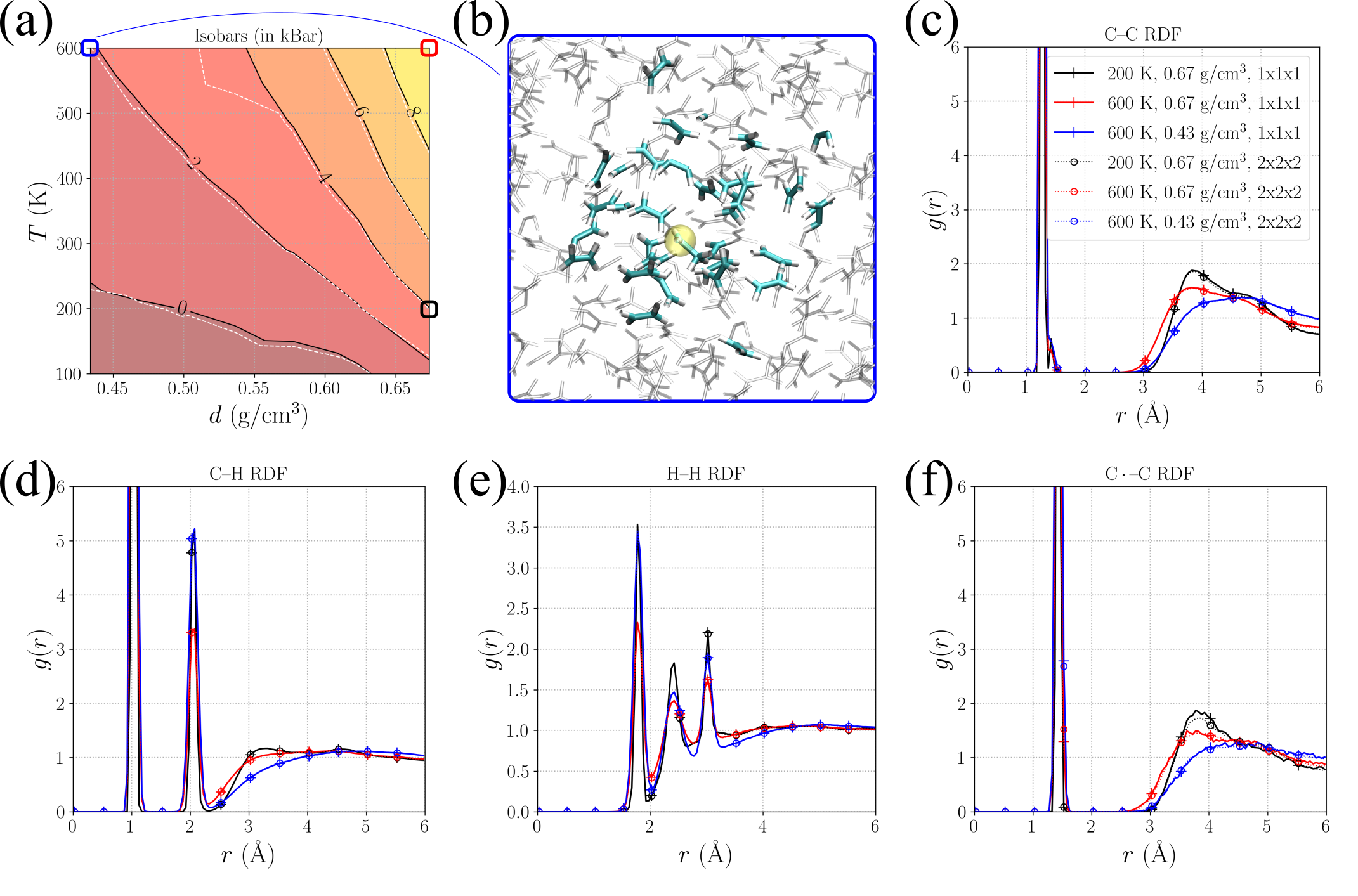}
    \caption{Assessment of finite-size effects on the solvated ethyl radical (1-mer). Panel (a) shows the isobars of \ce{(C2H5.)(C2H4)31} in a periodic unit cell (color-shaded contour lines labeled with isovalues in kbar) plotted against corresponding isobars initiated with a 2x2x2 supercell (white dashed lines). Panel (b) provides a graphical representation of the condensed-phase system with \ce{C} (\ce{H}) atoms in cyan (white) and the radical highlighted in yellow.
    Panels (c-f) shows the \ce{C}--\ce{C}, \ce{C}--\ce{H}, \ce{H}--\ce{H}, and \ce{C*}--\ce{C} radial distribution functions (RDFs) of three select thermodynamic conditions labeled as color-coded open squares in (a); the marker shapes denote the initial condition---thus also the system size---used in the DPMD simulations: crosses (dots) indicate a $1\times 1\times 1$ unit cell ($2\times 2\times 2$ supercell) as the starting configuration.
    }
    \label{fig:fse_1mer}
\end{figure*}

To characterize the oligomer radicals present in the high-pressure PE polymerization conditions, we first investigate the ethyl radical (i.e., initiated 1-mer).
We begin by assessing the finite-size effects of our 32-molecule system on thermodynamic properties based on the isobars over $T=100\mathrm{-}600$~K and $\rho=0.43\mathrm{-}0.66$~g/cm$^3$ with those obtained at a larger system size (a 256-molecule system initiated using a $2\times2\times2$ supercell as the starting configuration for each of these DPMD simulations).
The isobars are constructed by a collection of $200$-ps DPMD simulations ($100$-ps equilibration followed by $100$-ps production) for each combination of $T\in\{100,200,300,400,500,600\}$~K and $\rho\in\{0.43, 0.46, 0.49, 0.53, 0.57, 0.62, 0.66\}$~g/cm$^3$ with a timestep of $1.0$~fs with all \ce{H} replaced with \ce{D}.
As shown in Fig.~\ref{fig:fse_1mer}a, we found that the isobars sampled by the 32-molecule unit cells (illustrated in Fig.~\ref{fig:fse_1mer}b) are in good agreement with those from the 256-molecule supercells, indicating minimal finite-size effects in the sampled pressure over the observed range of $T$ and $\rho$.
Based on these isobars, the typical experimental high-pressure polymerization conditions ($T \geq 200$~$^\circ$C and $p \geq 500$~bar)~\cite{aggarwal_polyethylene_1957} approximately correspond to the region of $T \geq 473$~K and $\rho \geq 0.43$~g/cm$^3$.

In addition to the thermal properties, we also assessed finite-size effects on the structural properties using the \ce{C}--\ce{C}, \ce{C}--\ce{H}, \ce{H}--\ce{H}, and \ce{C*}--\ce{C} RDFs.
As shown in Fig.~\ref{fig:fse_1mer}c--f, we observed essentially identical average structure when comparing the 32-molecule and 256-molecule unit cells.
Based on this finite-size study, we will retain the 32-molecule unit cell to perform further analysis of other PE oligomers.
As expected, the strength of intramolecular correlations (e.g., distances less than $\approx$ 3~\AA{}) generally decreased with increasing temperature (from black to red curves) due to stronger thermal fluctuations. A decrease in density (from red to blue) reduces the condensed-phase effects (e.g., collision with neighboring molecules), leading to more uniform intramolecular structures (sharper peaks).
Intermolecular correlations (e.g., smooth plateau with an onset distances greater than $\approx$ 3~\AA{}) decreased upon increasing temperature (again due to stronger thermal fluctuations) and upon decreasing density (as the system approaches more gas-like behavior).
Interestingly, comparing Fig.~\ref{fig:fse_1mer}c and \ref{fig:fse_1mer}f, the intermolecular solvation structure of radical and non-radical carbon atoms appears nearly identical, indicating that the thermal (kinetic) energy at this experimental supercritical condition outweighs the differential binding energies between the radical and non-radical \ce{C} atoms.
Furthermore, we found that the radicals remain stable within the $100$-ps timescale of our DPMD simulations.
Specifically, we did not observe the association of two radicals during our DPMD simulation of the 256-molecule supercells, which contained 8 radicals total.
We observed that the radical pairs can approach each other to become first neighbors within a distance of $\approx 3$~\AA.
This observation is consistent with Flory's estimated mean duration of existence of a pair of radicals as first neighbors of $10^{-11}$s to $10^{-10}$s.~\cite{flory_principles_1953}

\begin{figure}[ht!]
    \centering
    \includegraphics[width=\linewidth]{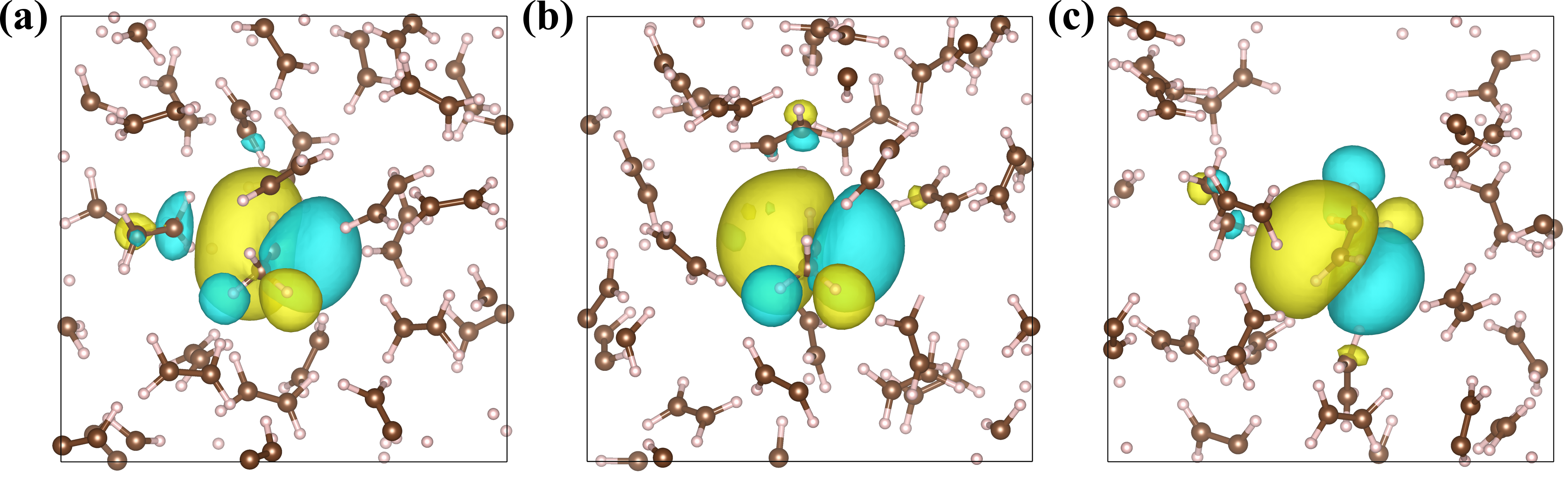}
    \caption{Three representative solvation structures (panels a--c) around an ethyl radical under a high-pressure polymerization of ethylene at $T=500$~K and $\rho=0.57$~g/cm$^3$.
    Graphical representation of the condensed-phase systems include: (1) ball-and-stick models for the nuclei with \ce{C} (\ce{H}) atoms in brown (white) spheres; (2) localized orbital isosurfaces for the radical electrons at isovalues of $\pm 0.8$~Bohr$^{-3/2}$ with positive (negative) lobes in yellow (cyan); and (3) bounding boxes with black edges for periodic unit cell used in the simulations.}
    \label{fig:radical_SCDM}
\end{figure}

We now analyze the equilibrated DPMD trajectories for the solvation structures around the ethyl radical at $T = 500$~K and $\rho = 0.57$~g/cm$^3$.
As shown Fig.~\ref{fig:radical_SCDM}, we found that the radical electrons can be visualized quite well with the isosurfaces of the SCDM orbital closest to the radical carbon with isovalues of $\pm 0.8$~Bohr$^{-3/2}$.
Based on this choice, we found that the ethyl radical electron can delocalize beyond the radical carbon to a few (1--3) neighboring ethylene molecules with ethylene-carbon-to-radical-carbon distance within $\approx 4.0$~\AA{}, indicating a fairly strong radical--$\pi$ interaction.
The $2\pm 1$ neighboring molecules that strongly interact with the radical can be understood by the approximate planar symmetry of the radical orbital (colloquially stated as a $p_{z}$ orbital in organic chemistry), leading to two equally possible directions for the radical to interact (from above or below the plane).

\section{Solvated PE Oligomer Radicals}

Having analyzed the ethyl radical, we now study PE oligomer radicals to explore how chain length affects solvation structure as polymerization proceeds.
We note in passing that the isobars demonstrate non-monotonic behavior as we extend the chain length of PE oligomers; a few factors could contribute to this oscillatory chain-length dependence, including (1) reduced effective molecular volume as monomers combine into oligomers, (2) increased molar fraction of the oligomer as the solvent monomers are added to the oligomer (a finite-size effect in our simulation), or (3) small MLFF inconsistencies due to lack of internal stress tensor information in the training data.
While a detailed investigation of these effects are outside the scope of this work, the consistent trends in $p(\rho,T)$ among the different oligomers already allow us to provide an approximate relationship (within a few kbar) between $p$ (which is the primary state point observable in experiments in conjunction with $T$) to $\rho$ (which is the more convenient state point variable for microscopic simulations).

\begin{figure}
    \centering
    \includegraphics[width=\linewidth]{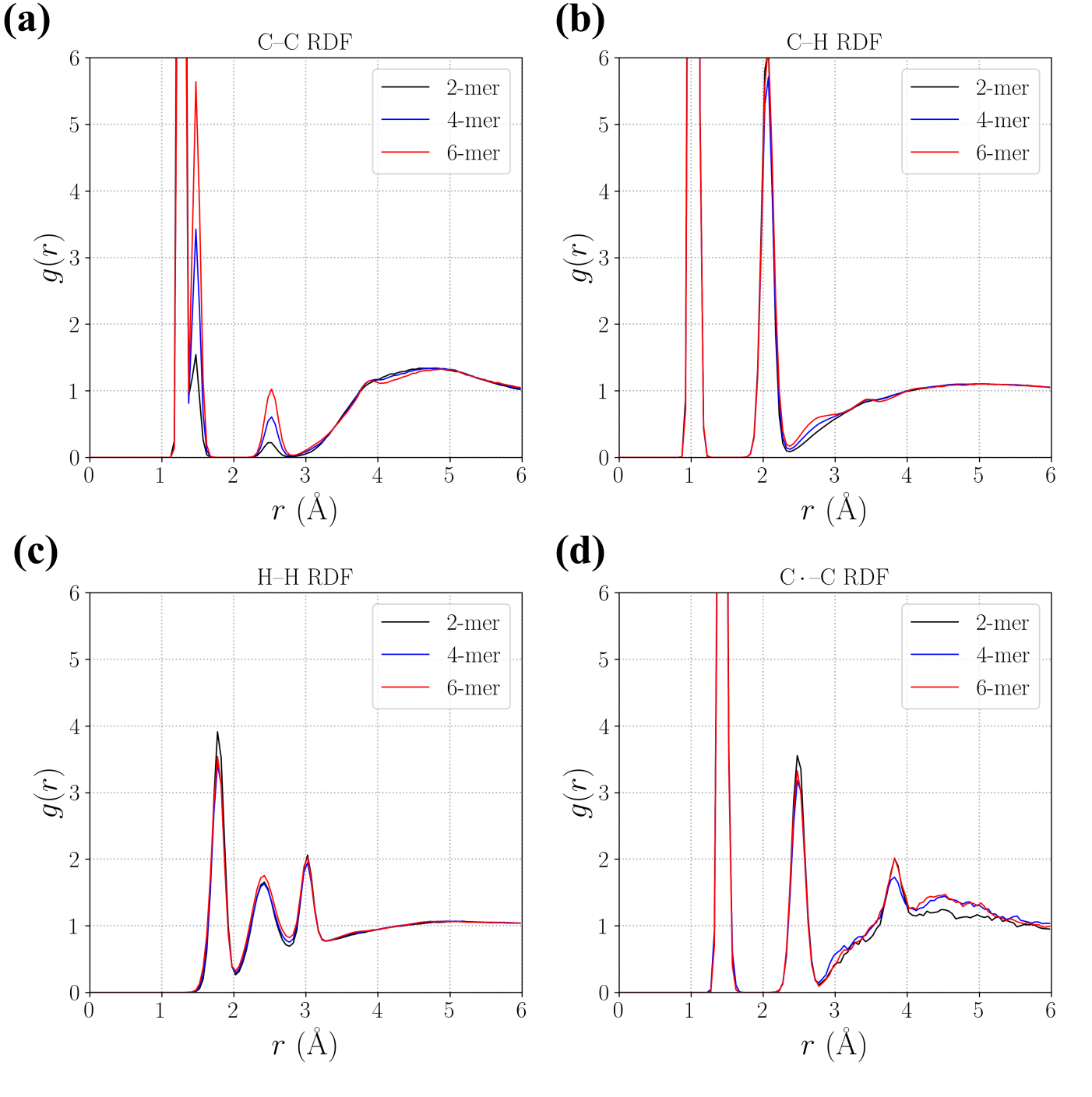}
    \caption{Chain-length-dependent radial distribution functions for (a) \ce{C}--\ce{C}, (b) \ce{C}--\ce{H}, (c) \ce{H}--\ce{H}, and (d) \ce{C*}--\ce{C} pairs sampled using the 32-molecule unit cell at $T=600$~K and $\rho=0.43$~g/cm$^3$.
    }
    \label{fig:rdf_nmers}
\end{figure}

To understand how the structural properties of the oligomer change as the chain length grows during polymerization, we investigate the RDFs of the oligomer-containing systems.
In Fig.~\ref{fig:rdf_nmers}, we observed that as the chain length increases, two signatures of developing intramolecular structure begin to appear in the \ce{C}--\ce{C} RDF: (1) the first peak has a growing shoulder representing the slightly longer \ce{C-C} single bond in the oligomer compared with the \ce{C=C} double bond in the monomer; (2) the second peak at around $r=2.5$~\AA{} represents the two \ce{C} atoms at $\beta$-position to each other (i.e., \ce{C} atoms separated via two connected \ce{C-C} bonds along the oligomer chain).
These intramolecular features grow as the chain length increases (correspondingly consuming the ethylene monomers).
The effect of chain growth is less visible in intramolecular contributions to \ce{C}--\ce{H} and \ce{H}--\ce{H} RDFs since these pair distances are typically mediated via indirect bonding along the oligomer chain and are smeared by intramolecular bond rotation and intermolecular contributions to the RDFs.
Interestingly, the \ce{C*}--\ce{C} RDF only exhibits minor differences with growing chain length, indicating that the solvation structure of the radical is similar across the short-chain alkanes.

\begin{figure}[ht!]
    \centering
    \includegraphics[width=\linewidth]{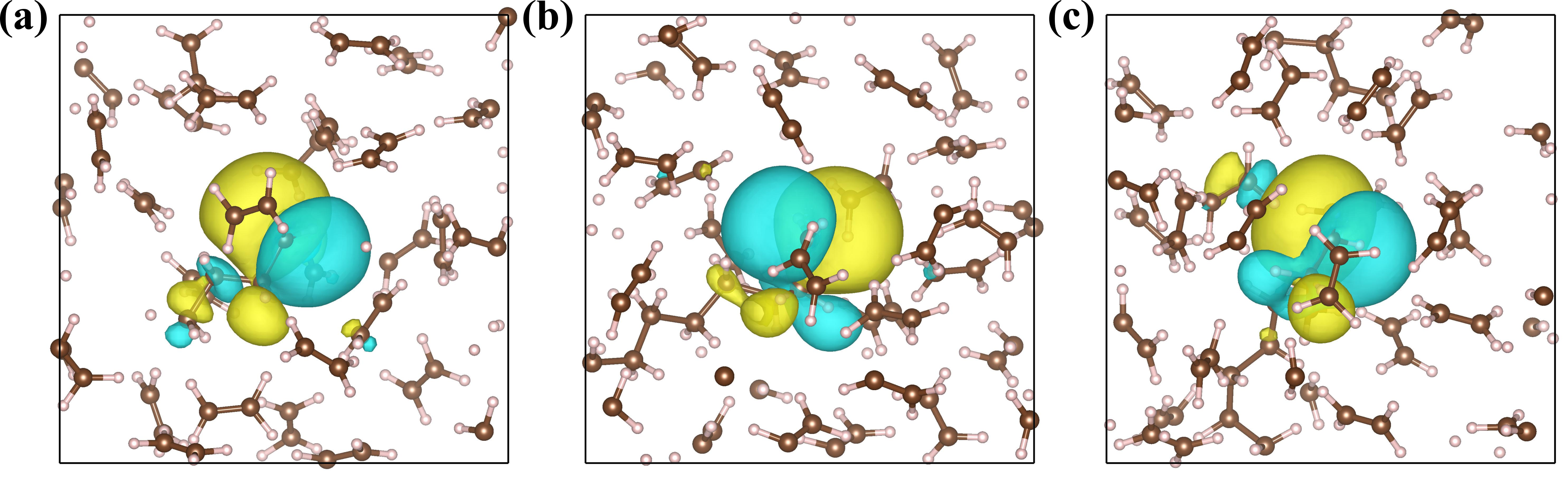}
    \caption{Representative solvation structures of PE oligomer radicals of lengths (a) $n=2$, (b) $n=4$, and (c) $n=6$ under a high-pressure polymerization of ethylene at $T=500$~K and $\rho=0.57$~g/cm$^3$ with the consistent graphical representation used in Fig.~\ref{fig:radical_SCDM}.
    }
    \label{fig:radical_2_4_6mer}
\end{figure}

Motivated by the convergent structural properties as functions of the chain length, we perform analogous localized orbital analysis as done in Sec.~\ref{sec:ethyl_rad} to further quantify the chain length dependence of the solvation structure around the oligomer radicals.
We consistently found that the PE oligomer radical electron (characterized with isovalues of $\pm 0.8$~Bohr$^{-3/2}$ in Fig.~\ref{fig:radical_2_4_6mer}) can again delocalize beyond the radical carbon to a few (1--3) neighboring ethylene molecules with ethylene-carbon-to-radical-carbon distance within $\approx 4.0$~\AA{}.
The rapidly converging solvation structure with respect to the chain length reveals the possibility of traversing the local chemical environments (i.e., the input needed by DP MLFFs) of a macromolecule using those from a small collection of oligomer models, thereby indicating the extensibility of our oligomer-trained DP MLFF to the PE polymers.
This rapidly converging local chemical environment around the reactive radical carbon is consistent with Flory's principle of equal reactivity, which states that the intrinsic reactivity of a functional group (here, the radical) approaches a constant quickly with respect to the polymer chain length~\cite{flory_principles_1953}.

As an aside, in practice, the local chemical environment of a solvated oligomer lacks the type of environments that involve interactions between segments of polymer chains.
However, this type of chain--chain interaction is present in our training data since, in this intentionally small periodic cell, the oligomer chain (e.g., $n=6$) can appear in the proximity of its periodic image(s).
Our initial tests with multi-chain simulations suggest that this procedure for generating training data may include sufficient configurations relevant to inter-segment interactions to extrapolate to PE melts or dense solutions---exploring this observation in detail will be the focus of a subsequent work.

\section{Extensibility of the Oligomer-Trained MLFF to Polymers}

\begin{figure*}[ht!]
    \centering
    \includegraphics[width=0.9\linewidth]{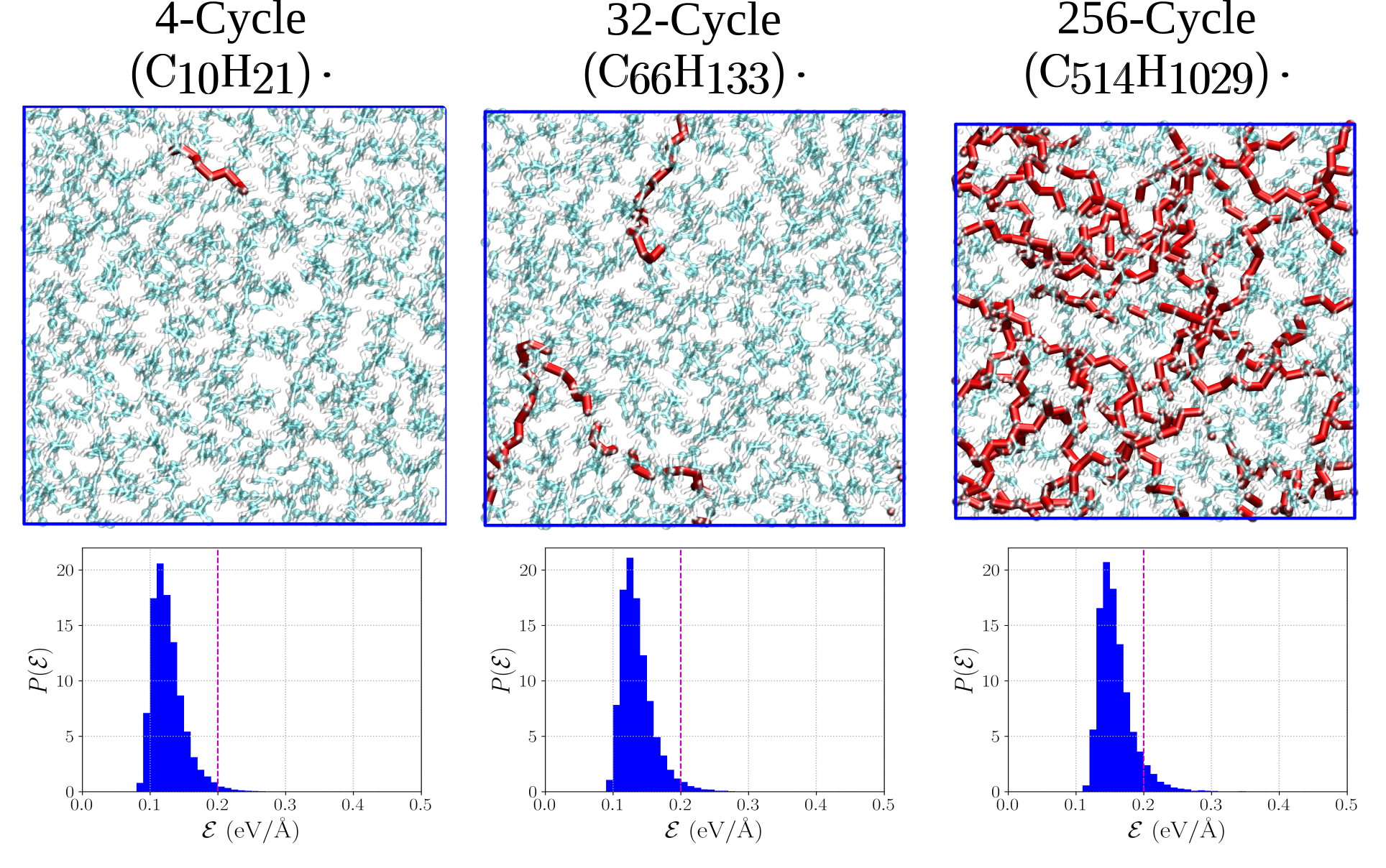}
    \caption{
    Stability of oligomer-trained MLFF for different PE chain lengths from $n=5$ (left column), $n=33$ (middle), and $n=257$ (right) within a $512$-molecule unit cell at $T=500$~K and $p=4$~kbar (pressure consistent with the initial liquid density as shown in Fig.~\ref{fig:fse_1mer}a). 
    For each column, the inset contains a representative snapshot of the DPMD simulation containing: (1) ball-and-stick models for the nuclei with \ce{C} (\ce{H}) atoms in translucent cyan (white) spheres; (2) blue bounding boxes for the periodic unit cells; and (3) red cylinders for polymer backbones.
    Beneath each inset is a model-deviation distribution plot to assess the precision of the DP MLFF, with the active-learning accuracy threshold labeled as a purple dashed line.
    }
    \label{fig:model_devi_evo_chain_length}
\end{figure*}

The convergence in the local chemical environment with growing oligomer chain length (Fig.~\ref{fig:rdf_nmers}) suggests that our short-ranged DP MLFF trained using oligomer data can be extensible to molecules with polymeric lengths.
To test this hypothesis, we used the oligomer-trained MLFF to perform 1-ns DPMD simulations for each of the following systems: a short oligomer from 4 chain-growth cycles ($n=5$, \ce{C10H21*}, $141$~g/mol), a long oligomer from 32 cycles ($n=33$, \ce{C66H133*}, $927$~g/mol), and a polymer from 256 cycles ($n=257$, \ce{C514H1029*}, $7,211$~g/mol); in this case, we sample the $NpT$ ensemble using the equations of motion of Shinoda and coworkers~\cite{shinoda_rapid_2004} with a relaxation time of $1.0$~ps at $T=500$~K and $p=4$~kbar to account for the natural reduction of pressure as the polymerization process replaces (larger-volume) intermolecular vdW contact with (smaller-volume) \ce{C-C} bonds. 
As shown in Fig.~\ref{fig:model_devi_evo_chain_length}, the model deviation distributions (which we used as a proxy for the accuracy/reliability of the DP MLFF) resulting from these oligomer and polymer simulations are essentially independent of chain length, directly supporting the extensibility of the oligomer-trained MLFF to macromolecules. All systems studied remained stable for the duration of the 1-ns DPMD simulations.

\begin{figure*}[ht!]
    \centering
    \includegraphics[trim={80 00 80 10}, clip,width=0.9\linewidth]{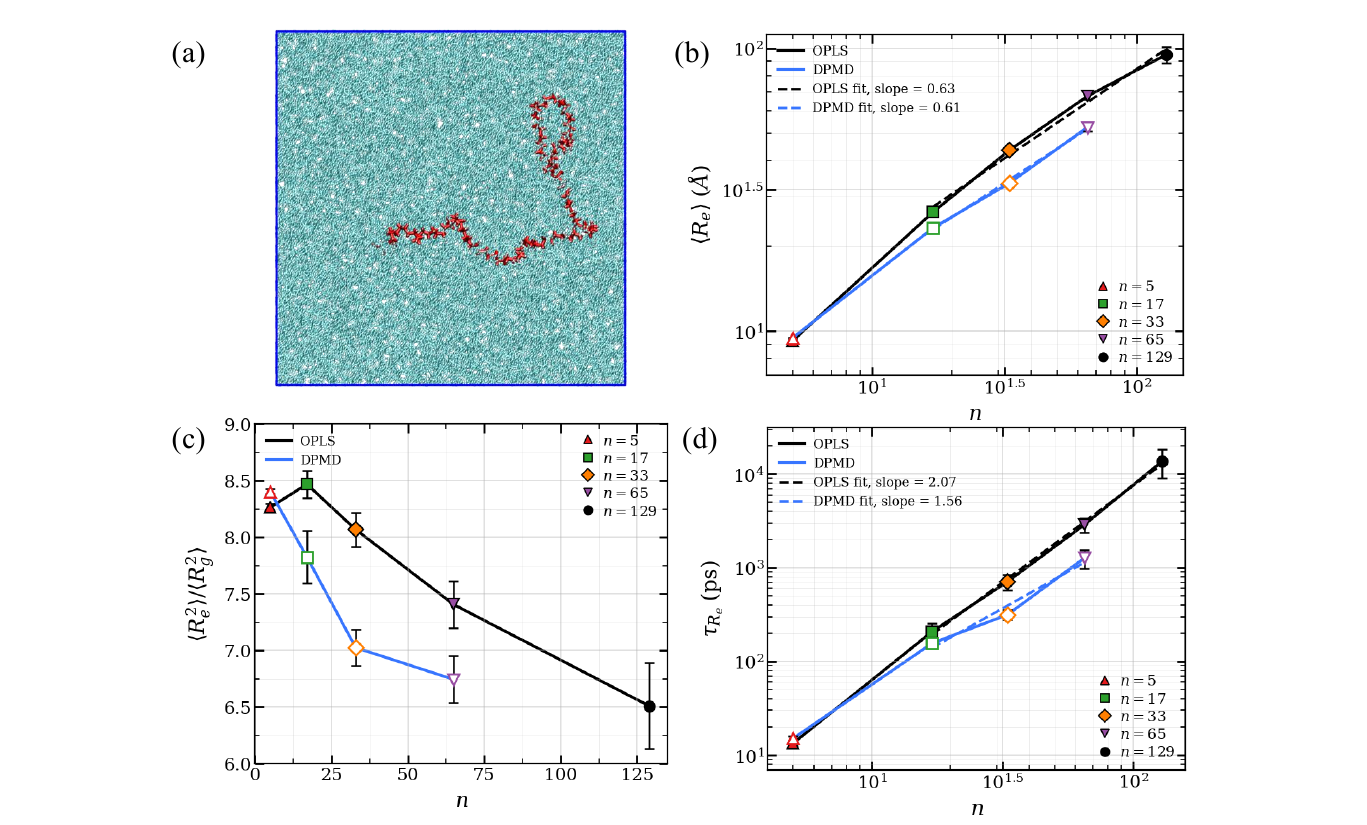}
    \caption{
    Single-chain statistics. Panel (a) shows a representative simulation snapshot of the DPMD simulation for the $n=65$ carbon chain. Panel (b) shows the scaling of chain end-to-end distance, $\braket{R_e}$, as a function of degree of polymerization $n$, panel (c) shows the ratio $\braket{R_e^2}/\braket{R_g^2}$, where $R_g$ is the radius of gyration. Panel (d) describes chain dynamics via the relaxation time in $R_e$, $\tau_{R_e}$. In all panels, black lines denote the L-OPLS force field, blue lines denote DPMD results, and error bars denote $95\%$ confidence intervals.
    }
    \label{fig:Chain_scaling}
\end{figure*}

\section{Macromolecular Structure and Dynamics from First Principles}

Given the promising results regarding the stability and transferability of our DP MLFF across different chain lengths, we now analyze the structural and dynamic properties of a single solvated PE chain with varying molecular weight.
We performed extensive $NVT$ simulations at $T=500$~K and $\rho=0.56$~g/cm$^3$ (characteristic of supercritical ethylene) using DPMD, and compared our first-principles results to a variant of the widely used classical atomistic force field, OPLS~\cite{jorgensen_development_1996}, optimized for its predictive accuracy for long-chain alkanes (L-OPLS)~\cite{siu_optimization_2012}. In contrast to the intentionally small unit cells used for model training, in these simulations we ensured that all systems were large enough to prevent any unphysical intra-chain interactions through the periodic boundaries such that our results were reflective of single-chain statistics in the dilute limit. As a result, our DPMD simulations included up to $43,904$ total atoms (Fig.~\ref{fig:Chain_scaling}a) for the longest chain length considered, and our combined DPMD sampling time totaled more than $350$~ns. We note that these chain scaling simulations did not contain a carbon radical; these simulations were stable across all conditions explored despite the training data being focused on radical-containing systems. More details about the chain scaling simulations can be founding Appendix~\ref{app:chain-scaling}.

Fig.~\ref{fig:Chain_scaling} summarizes the scaling of the equilibrium chain dimensions and chain relaxation dynamics as a function of the number of ethlyene monomers in the alkane chain, $n$ (i.e., half the number of carbons). The mean end-to-end distance (Fig.~\ref{fig:Chain_scaling}b) follows $\braket{R_e} \sim n^v$, yielding $v=0.63$ for L-OPLS and $0.61$ for the PBE0+D3 DP MLFF, close to the expected value for excluded-volume chains in a good solvent ($v\approx0.5877$)~\cite{li_critical_1995}. We omitted $n=5$ from the scaling regression because low-molecular-weight alkane chains assume elongated configurations, with their dimensions strongly influenced by short-range conformational constraints~\cite{mondello_dynamics_1998,jeong_mass_2015}. The DP MLFF predicts somewhat smaller chain dimensions than L-OPLS, which could reflect a more flexible chain or slightly less favorable polymer-solvent interactions. Short-chain effects are also evident in the ratio $\braket{R_e^2}/\braket{R_g^2}$ (Fig.~\ref{fig:Chain_scaling}c), where $R_g$ is the radius of gyration. For both models, $\braket{R_e^2}/\braket{R_g^2}$ decreases with increasing chain length toward the good-solvent limit of approximately 6.25~\cite{li_critical_1995,sherck_end--end_2020}. The decrease in this ratio as $n$ increases reflects the diminishing influence of local backbone stiffness on the chain conformation, as longer chains progressively lose their short-chain elongation and approach coil-like excluded-volume statistics~\cite{baschnagel_monte_1992}. The lower $\braket{R_e^2}/\braket{R_g^2}$ ratios obtained with the DP MLFF at intermediate chain lengths similarly indicate an earlier onset of flexible behavior with increasing $n$, whereas L-OPLS retains a more extended conformation.

For both the PBE0+D3 DP MLFF and L-OPLS, the scaling of the chain relaxation time, $\tau_{R_e}$ (Fig. \ref{fig:Chain_scaling}d, defined via the decay of the chain $R_e$ autocorrelation function) is in a similar range to the prediction of standard models such as Zimm ($\approx 1.8$)~\cite{rubinstein_polymer_2003}. The chain relaxation scaling exponent is strongly impacted by statistical sampling, and despite our extensive simulations, the accuracy of our estimates is somewhat limited by the timescales accessible to DPMD. At comparable $n$, the DP MLFF generally exhibits shorter characteristic times than L-OPLS, suggesting faster reorientation of the end-to-end vector. Given the limited temporal window sampled by DPMD, however, the extent of this difference remains only qualitatively resolved. Overall, both models exhibit reasonable polymer-like scaling behavior, characteristic of good-solvent conditions. The DP MLFF appears to predict a slightly more flexible chain with a faster relaxation time than L-OPLS, which could be due to additional conformational flexibility (anharmonicity) afforded by the lack of any pre-supposed functional forms for the intramolecular bonds, angles, or dihedrals. While supercritical ethylene is widely used as a polymerization solvent and has been shown to dissolve short-chain polyethylene waxes~\cite{heukelbach_critical_1998}, detailed experimental characterization of polymer chain dimensions in supercritical solvents is challenging due to the high temperatures and pressures required. Thus, our simulations provide a predictive first-principles window into PE chain scaling and dynamics under these complex and industrially relevant conditions.

\section{Conclusions and Future Work}

In this work, we developed a systematic active machine learning workflow to train a DP MLFF with high chemical realism from first-principles electronic structure theory at the PBE0+D3 hybrid DFT for polyethylene based on a dataset of monomer-solvated small oligomers in the condensed phase.
We assessed the thermodynamic properties and solvation structures of PE oligomer radicals using DPMD simulations.
Our observation of the rapid convergence of both the nuclear and electronic solvation environments around the radical site with respect to chain length reveals the extensibility of the oligomer-trained DP MLFF to the macromolecular limit and aligns well with the longstanding principle of equal reactivity in polymer chemistry.
The extensibility of the DP MLFF is confirmed by direct DPMD simulations of macromolecular PE with chain length up to $n=257$.
While not explicitly trained, polymer chain--chain interactions are included in the training data through oligomer interactions with their periodic images, due to the finite-size effect of intentionally small periodic unit cells.
We then compared the trained PBE0+D3 DP MLFF with the conventional L-OPLS force field and found that our DP model captured realistic single-chain scaling behavior in terms of both structure and dynamics. Our simulations confirm that supercritical ethylene behaves as a good solvent for short-chain PE molecules, providing a first-principles look into polymer physics properties for these ubiquitous yet challenging-to-study conditions.
Taken together, this work provides new insights into polymer behavior from the level of electronic structure to the macromolecular polymer chain and beyond, in a single, first-principles-derived modeling framework. Our approach helps pave the way towards bridging the chemical realism of first-principles electronic structure theory and predictions of large-scale polymer physical properties.

For future applications, the simple training workflow we developed using small oligomers along with an intentionally small unit cell to account for chain--chain interactions can systematically be applied to other macromolecules.
Our approach also offers a pathway for systematically understanding the interactions associated with complex polymer systems when comparing MLFFs trained with different DFT approximations; for instance, the effect of dispersion forces by comparing predictions with and without vdW corrections as done in Ref.~\onlinecite{hong_first-principles-based_2021}. We anticipate that these models can be extended to directly study reactive processes such as polymer chain synthesis for bottom-up polymer structure and property prediction, as well as studies of polymer upcycling, recycling, and degradation. We also anticipate possible applications in studying charged and ion-containing polymers, polymer membranes, and advanced electronic materials, where capturing the interplay between electronic structure and polymer behavior will be critical. Overall, this approach will enable key future fundamental and applied studies across the breadth of computational polymer science.

\section*{Acknowledgments}

This work was supported by the University of North Texas (UNT) and Lehigh University.
We acknowledge computational resources from the UNT CASCaM HPC cluster via National Science Foundation OAC-2117247, the Texas Advanced Computing Center (TACC) at The University of Texas at Austin, and Lehigh University's Research Computing infrastructure partially supported by NSF Award 2019035.

\appendix

\section{Computational Details \label{app:compt_details}}

\subsection{Electronic Structure Calculations}
%
All electronic structure calculations for the DP MLFF training data generation were performed using the \texttt{SeA} high-throughput hybrid DFT engine~\cite{ko_high-throughput_2023} with the Grimme DFT-D3 dispersion correction~\cite{grimme_consistent_2010} in Quantum ESPRESSO~\cite{giannozzi_quantum_2009,giannozzi_advanced_2017,giannozzi_quantum_2020}.
During each of these calculations, we used a simple cubic cell containing \ce{C64H129*} with periodic boundary conditions. 
Optimized norm-conserving Vanderbilt (ONCV) pseudopotentials~\cite{hamann_optimized_2013} from the SG15 collection~\cite{schlipf_optimization_2015} are used to model the interactions between the valence electrons and ions (which include nuclei and their frozen-core electrons).
We used a planewave basis with a maximum kinetic energy of $120$~Ry to represent the pseudo-wavefunctions for the valence electrons.
The Brillouin zone was sampled only at the gamma point.
We used an SCF convergence threshold of $10^{-6}$~Ry.
For all \texttt{SeA} calculations, we used the default setting of the system-independent orbital coverage threshold ($\epsilon=10^{-3.5}$)~\cite{ko_high-throughput_2023}.

\subsection{Initial Data Generation}

To generate the initial training data, we perform AIMD simulations of liquid ethylene doped with an ethyl radical (which could also be considered a PE oligomer of length $n = 1$) using a \texttt{SeA}-enabled \texttt{PWscf}~\cite{ko_high-throughput_2023} package in \texttt{Quantum ESPRESSO}~\cite{giannozzi_quantum_2009,giannozzi_advanced_2017,giannozzi_quantum_2020}.
To do so, we select a starting density using the experimental liquid ethylene density of $\rho = 0.57$~g/cm$^3$ at around ($170$~K, $1$~bar) and use the \texttt{PACKMOL} software package~\cite{martinez_packmol_2009} to generate a proto-configuration containing $31$ ethylene molecules and one ethyl radical.
To sample a broad range of pressures, we uniformly scale the above proto-configuration along each linear dimension by a factor $s\in\{0.95, 1.00, 1.05, 1.10\}$ to prepare four initial configurations with volume scalings of $\{86\%, 100\%, 116\%, 133\%\}$ (leading to densities of $\{0.66, 0.57, 0.49, 0.43\}$~g/cm$^3$).
For each of the four initial configurations, three AIMD simulations at the PBE0+D3 level were performed within the canonical $NVT$ ensemble using the stochastic velocity rescaling thermostat (SVR, timescale of $20$~fs)~\cite{bussi_canonical_2007} for $\approx 1$~ps at $T=200$~K, $400$~K, and $600$~K.
In these AIMD simulations, we opt to simulate fully deuterated molecules to slow down the \ce{C-H} vibration and use a longer-than-typical timestep of $\approx 2$~fs to allow faster (approximated) initial exploration to a set of $5,400$ data points from all combinations of densities and temperatures (saving every single time frame in the dataset).
Because we are propagating the MD equations of motion using classical dynamics, the resulting structural properties and the relationship between atomic coordinates and potential energies/forces are unaffected by this isotope substitution (\ce{H}$\rightarrow$\ce{D}).

To build a PE oligomer of length $n=2$ (i.e., a 1-butyl radical), we start from the proto-configurations described above ($n=1$, four densities) and perform constrained AIMD to gradually reduce the distance between the radical C in the ethyl radical and another randomly selected \ce{C} atom in a neighboring ethylene molecule to obtain four proto-configurations of $n=2$ (one for each density).
Starting from the $n=2$ configurations, we again perform unbiased $NVT$ AIMD data generation for all $12$ combinations of three temperatures and four densities to collect another set of $5,400$ data points.
This process is repeated until $n=6$, leading to an initial training set of $5,400\times 6 = 32,400$ PBE0+D3 energy and force data points.

\subsection{Active Learning}

Starting from a DP MLFF trained on the initial dataset, we perform active machine learning~\cite{zhang_active_2019} using the \texttt{DP-Gen} platform~\cite{zhang_dp-gen_2020} for an integrated 3-step workflow control (Fig.~\ref{fig:ml_scheme}).
For individual steps of the active learning cycle, we first used \texttt{DeepMD-kit}~\cite{wang_deepmd-kit_2018,zeng_deepmd-kit_2023} to train a DP model, then \texttt{LAMMPS}~\cite{plimpton_fast_1995,thompson_lammps_2022} to run MD simulations using the trained DP, and finally a \texttt{SeA}-enabled version~\cite{ko_high-throughput_2023} of \texttt{Quantum ESPRESSO}~\cite{giannozzi_quantum_2009,giannozzi_advanced_2017,giannozzi_quantum_2020} to perform first-principles electronic structure labeling of selected configurations (i.e., candidates) to be incorporated in the data set. In order to train the DP MLFF, we used \texttt{DeepMD-kit} settings as described in Ref.~\onlinecite{ko_high-throughput_2023}.
To account for our larger data size, we increased the number of training steps to $1$M for the initial active learning stage (Stages 1-3 in Table~\ref{tab:al_protocol}) and $2$M in the refinement stage (Stages 4-5 in Table~\ref{tab:al_protocol}).

To prepare the initial configurations for the MD exploration of each oligomer, we collected one structure from each of the $12$ AIMD trajectories with different thermodynamic conditions.
Starting from these $12$ initial configurations, we performed DPMD simulations within the $NVT$ ensemble and gradually (stepwise) increase the temperature in subsequent active learning cycles (after convergence) from $200$~K, to $400$~K, and then to $600$~K.

\begin{table}[ht!]
    \centering
    \begin{tabular}{c c c c}
\hline
\hline
\,Stage\,& \,$T$ (K)\,   & New\,Data Size\, & \,$\%\text{acc}$\, \\
\hline
AIMD     & [200,400,600] & 32,400             & -- \\
1        & 200           & \phantom{0}7,253	  & 99\%  \\
2        & 400           & \phantom{0}6,217	  & 99\%  \\
3        & 600           & 31,390             & 88\%--98\% \\
4        & 600           & 15,545             & 96\% \\
5        & 600           & \phantom{0}6,269	  & 98\% \\
\hline
\multicolumn{4}{l}{Total: $99,074$ PBE0+D3 energy and forces}\\
\hline
\hline
    \end{tabular}
    \caption{Summary of active-learning protocol for PE oligomers from $n=1\mathrm{-}6$.
    The first column shows the coarse-grained active learning stages, starting from (1) the initial data generation with AIMD, (2) the first round of active learning (Stages 1--3) in which we explore each oligomer with a gradually increasing temperatures, and then (3) the second round of active learning cycles (Stages 4--5) in which we combine data from all oligomers and used refined training settings.
    The second, third, and fourth columns show the temperatures, new data size, and accuracy ratio $\%\text{acc}$, which is defined based on the fraction of configurations with model deviation below $0.20$~eV/\AA{}.
    Based on this protocol, we obtained an ensemble of four DP MLFFs with $\%\text{acc}\geq 98\%$ across our target thermodynamic range.
    For each step, we consistently explore system densities of $\rho\in\{0.43, 0.49, 0.57, 0.66\}$~g/cm$^3$.
}
    \label{tab:al_protocol}
\end{table}
To identify candidate structures for first-principles relabeling, we use the conventional model deviation $\mathcal{E}=\max_I \sqrt{\braket{|| \bm {F}_I - \braket{\bm {F}_I} ||}}$ (the maximum standard deviation of the predicted atomic forces among an ensemble of four equivalent DP models trained with different random seeds)~\cite{zhang_active_2019}, in which $\bm {F}_I$ denotes the atomic force on the $I$-th atom and $\braket{}$ denotes the ensemble average over equivalent DP models.
Specifically, we collect a candidate structure when its $\mathcal{E}$ value falls in the window of $[0.20,0.35]$~eV/\AA{}, similar to the settings used in a previous study of the water phase diagram~\cite{zhang_phase_2021}.
The active learning process leads to $99,074$ PBE0+D3 data points (Table~\ref{tab:al_protocol}) for PE oligomers spanning $n=1 \mathrm{-} 6$.
We stress that while four equivalent DP MLFFs are trained and used to calculate the model deviation metric used to identify new candidate structures for labeling, after the model development process is finished only one of those models is used to propagate further DPMD simulations of PE oligomers and polymers.
We note in passing that due to the growing data size, the accuracy level becomes unstable in Stage 3 of Table~\ref{tab:al_protocol}.
Therefore, we had to use the refinement setting ($2$M training steps) to achieve a stable DP MLFF in Stages 4--5 of Table~\ref{tab:al_protocol}.
A typically converged model deviation profile can be found in the lower-left panel of Fig.~\ref{fig:model_devi_evo_chain_length}.

\section{Chain Scaling Simulations}\label{app:chain-scaling}

To evaluate the PE chain scaling statistics, MD simulations of single PE chains solvated in ethylene were performed in \texttt{LAMMPS}~\cite{thompson_lammps_2022} using the DP MLFF developed herein and the classical atomistic L-OPLS force field~\cite{siu_optimization_2012}, which was optimized for long chain alkanes. PE degrees of polymerization $n=5$, $n=17$, $n=33$, and $n=65$ were evaluated for both models, with L-OPLS simulations extended to $n=129$, and the respective simulation cells contained 632, 4,640, 20,720, 43,904, and 72,116 total atoms. The system sizes were chosen to avoid any possibility of intrachain interactions through the periodic boundary. In the L-OPLS model, the intramolecular potential includes harmonic bond stretching and angle bending, along with an explicit torsional contribution governing rotation about the polymer backbone. Nonbonded interactions were represented by 12–6 Lennard–Jones and Coulombic terms, with parameters assigned by atom type and cross-interactions determined via geometric mixing rules. Long-range electrostatics were evaluated using the particle–particle particle–mesh (PPPM) solver with a relative force accuracy of $10^{-4}$.

Simulations to evaluate chain scaling were conducted in the canonical ($NVT$) ensemble at $T=500$~K and $\rho=0.56$~g/cm$^3$ using Nos\'e–Hoover thermostat with a \texttt{LAMMPS} thermostat damping parameter $\tau_{damp}=100\ fs$, the \texttt{LAMMPS} default Nos\'e–Hoover chain with length 3, and an integration time step of $\Delta t =1\ fs$.  Initial configurations were prepared with \texttt{moltemplate}~\cite{jewett_moltemplate_2021}, and then subjected to energy minimization to relieve steric overlaps, followed by a 5 ns equilibration period using the L-OPLS force field to allow the chain conformations to relax. For the DPMD simulations, we then transitioned the interatomic potential to our trained DP MLFF, and discarded the subsequent 50 ps to eliminate transient artifacts associated with the force-field switch. L-OPLS simulations continued directly to the sampling stage. Structural relaxation was quantified via the end-to-end unit vector autocorrelation function, $C(t)=\braket{\bm{u}(t)\cdot\bm{u}(0)}$, where $\braket{\ }$ denotes averaging over all recorded configurations~\cite{foteinopoulou_structure_2008}. As the chains relax, $C(t)$ decays from unity to zero as the polymer loses memory of its original conformation, where a steeper decay reflects faster structural decorrelation.
The characteristic chain relaxation time $\tau_{R_e}$ was defined at the intersection point where $C(\tau_{R_e})=1/e$. 
Production runs were subsequently carried out for a total duration of $10\tau_{R_e}$ per replica, with a minimum of 10 independent replicas per chain. Uncertainties for all properties were estimated via $95\%$ confidence intervals calculated from the standard error across 30 block averages from the simulation trajectories.

\bibliography{references}

@article{ko_high-throughput_2023,
	title = {High-{Throughput} {Condensed}-{Phase} {Hybrid} {Density} {Functional} {Theory} for {Large}-{Scale} {Finite}-{Gap} {Systems}: {The} {SeA} {Approach}},
	volume = {19},
	doi = {10.1021/acs.jctc.2c00827},
	number = {13},
	journal = {J. Chem. Theory Comput.},
	author = {Ko, Hsin-Yu and Calegari Andrade, Marcos F. and Sparrow, Zachary M. and Zhang, Ju-an and DiStasio Jr., Robert A.},
	year = {2023},
	pages = {4182--4201},
}

@article{zeng_deepmd-kit_2023,
	title = {{DeePMD}-kit v2: {A} software package for deep potential models},
	volume = {159},
	issn = {0021-9606},
	shorttitle = {{DeePMD}-kit v2},
	url = {https://doi.org/10.1063/5.0155600},
	doi = {10.1063/5.0155600},
	number = {5},
	urldate = {2024-08-28},
	journal = {J. Chem. Phys.},
	author = {Zeng, Jinzhe and Zhang, Duo and Lu, Denghui and Mo, Pinghui and Li, Zeyu and Chen, Yixiao and Rynik, Mari{\'a}n and Huang, Li{\textquoteright}ang and Li, Ziyao and Shi, Shaochen and Wang, Yingze and Ye, Haotian and Tuo, Ping and Yang, Jiabin and Ding, Ye and Li, Yifan and Tisi, Davide and Zeng, Qiyu and Bao, Han and Xia, Yu and Huang, Jiameng and Muraoka, Koki and Wang, Yibo and Chang, Junhan and Yuan, Fengbo and Bore, Sigbj{\o}rn L{\o}land and Cai, Chun and Lin, Yinnian and Wang, Bo and Xu, Jiayan and Zhu, Jia-Xin and Luo, Chenxing and Zhang, Yuzhi and Goodall, Rhys E. A. and Liang, Wenshuo and Singh, Anurag Kumar and Yao, Sikai and Zhang, Jingchao and Wentzcovitch, Renata and Han, Jiequn and Liu, Jie and Jia, Weile and York, Darrin M. and E, Weinan and Car, Roberto and Zhang, Linfeng and Wang, Han},
	month = aug,
	year = {2023},
	pages = {054801},
}

@article{wang_deepmd-kit_2018,
	title = {{DeePMD}-kit: {A} deep learning package for many-body potential energy representation and molecular dynamics},
	volume = {228},
	issn = {0010-4655},
	shorttitle = {{DeePMD}-kit},
	url = {https://www.sciencedirect.com/science/article/pii/S0010465518300882},
	doi = {10.1016/j.cpc.2018.03.016},
	urldate = {2019-01-04},
	journal = {Comput. Phys. Commun.},
	author = {Wang, Han and Zhang, Linfeng and Han, Jiequn and E, Weinan},
	month = jul,
	year = {2018},
	pages = {178--184},
}

@article{zhang_dp-gen_2020,
	title = {{DP}-{GEN}: {A} concurrent learning platform for the generation of reliable deep learning based potential energy models},
	volume = {253},
	issn = {0010-4655},
	shorttitle = {{DP}-{GEN}},
	url = {http://www.sciencedirect.com/science/article/pii/S001046552030045X},
	doi = {10.1016/j.cpc.2020.107206},
	urldate = {2020-07-17},
	journal = {Comput. Phys. Commun.},
	author = {Zhang, Yuzhi and Wang, Haidi and Chen, Weijie and Zeng, Jinzhe and Zhang, Linfeng and Wang, Han and E, Weinan},
	month = aug,
	year = {2020},
	pages = {107206},
}

@article{zhang_active_2019,
	title = {Active learning of uniformly accurate interatomic potentials for materials simulation},
	volume = {3},
	url = {https://link.aps.org/doi/10.1103/PhysRevMaterials.3.023804},
	doi = {10.1103/PhysRevMaterials.3.023804},
	number = {2},
	urldate = {2019-10-07},
	journal = {Phys. Rev. Materials},
	author = {Zhang, Linfeng and Lin, De-Ye and Wang, Han and Car, Roberto and E, Weinan},
	month = feb,
	year = {2019},
	pages = {023804},
}

@incollection{zhang_end--end_2018,
	address = {Red Hook},
	title = {End-to-end {Symmetry} {Preserving} {Inter}-atomic {Potential} {Energy} {Model} for {Finite} and {Extended} {Systems}},
	url = {http://papers.nips.cc/paper/7696-end-to-end-symmetry-preserving-inter-atomic-potential-energy-model-for-finite-and-extended-systems.pdf},
	urldate = {2019-03-22},
	booktitle = {Advances in {Neural} {Information} {Processing} {Systems} 31},
	publisher = {Curran Associates},
	author = {Zhang, Linfeng and Han, Jiequn and Wang, Han and Saidi, Wissam and Car, Roberto and E, Weinan},
	editor = {Bengio, S. and Wallach, H. and Larochelle, H. and Grauman, K. and Cesa-Bianchi, N. and Garnett, R.},
	year = {2018},
	pages = {4436--4446},
}

@article{zhang_deep_2018,
	title = {Deep {Potential} {Molecular} {Dynamics}: {A} {Scalable} {Model} with the {Accuracy} of {Quantum} {Mechanics}},
	volume = {120},
	shorttitle = {Deep {Potential} {Molecular} {Dynamics}},
	url = {https://link.aps.org/doi/10.1103/PhysRevLett.120.143001},
	doi = {10.1103/PhysRevLett.120.143001},
	number = {14},
	urldate = {2019-01-03},
	journal = {Phys. Rev. Lett.},
	author = {Zhang, Linfeng and Han, Jiequn and Wang, Han and Car, Roberto and E, Weinan},
	year = {2018},
	pages = {143001},
}

@article{becke_density-functional_1993,
	title = {Density-{Functional} {Thermochemistry}. {III}. {The} {Role} of {Exact} {Exchange}},
	volume = {98},
	url = {http://scitation.aip.org/content/aip/journal/jcp/98/7/10.1063/1.464913},
	doi = {10.1063/1.464913},
	number = {7},
	journal = {J. Chem. Phys.},
	author = {Becke, Axel D.},
	month = apr,
	year = {1993},
	pages = {5648--5652},
}

@article{ko_enabling_2021,
	title = {Enabling {Large}-{Scale} {Condensed}-{Phase} {Hybrid} {Density} {Functional} {Theory}-{Based} {Ab} {Initio} {Molecular} {Dynamics} {II}: {Extensions} to the {Isobaric}{\textendash}{Isoenthalpic} and {Isobaric}{\textendash}{Isothermal} {Ensembles}},
	volume = {17},
	issn = {1549-9618},
	shorttitle = {Enabling {Large}-{Scale} {Condensed}-{Phase} {Hybrid} {Density} {Functional} {Theory}-{Based} {Ab} {Initio} {Molecular} {Dynamics} {II}},
	url = {https://doi.org/10.1021/acs.jctc.0c01194},
	doi = {10.1021/acs.jctc.0c01194},
	number = {12},
	urldate = {2021-12-14},
	journal = {J. Chem. Theory Comput.},
	author = {Ko, Hsin-Yu and Santra, Biswajit and DiStasio, Robert A.},
	month = dec,
	year = {2021},
	pages = {7789--7813},
}

@article{ko_enabling_2020,
	title = {Enabling {Large}-{Scale} {Condensed}-{Phase} {Hybrid} {Density} {Functional} {Theory} {Based} {Ab} {Initio} {Molecular} {Dynamics}. 1. {Theory}, {Algorithm}, and {Performance}},
	volume = {16},
	issn = {1549-9618},
	url = {https://doi.org/10.1021/acs.jctc.9b01167},
	doi = {10.1021/acs.jctc.9b01167},
	number = {6},
	urldate = {2020-06-10},
	journal = {J. Chem. Theory Comput.},
	author = {Ko, Hsin-Yu and Jia, Junteng and Santra, Biswajit and Wu, Xifan and Car, Roberto and DiStasio Jr., Robert A.},
	month = jun,
	year = {2020},
	pages = {3757--3785},
}

@article{ko_isotope_2019,
	title = {Isotope effects in liquid water via deep potential molecular dynamics},
	volume = {117},
	issn = {0026-8976},
	url = {https://doi.org/10.1080/00268976.2019.1652366},
	doi = {10.1080/00268976.2019.1652366},
	number = {22},
	urldate = {2019-11-26},
	journal = {Mol. Phys.},
	author = {Ko, Hsin-Yu and Zhang, Linfeng and Santra, Biswajit and Wang, Han and E, Weinan and DiStasio Jr, Robert A. and Car, Roberto},
	month = nov,
	year = {2019},
	pages = {3269--3281},
}

@article{adamo_toward_1999,
	title = {Toward reliable density functional methods without adjustable parameters: {The} {PBE0} model},
	volume = {110},
	issn = {00219606},
	shorttitle = {Toward reliable density functional methods without adjustable parameters},
	url = {http://jcp.aip.org/resource/1/jcpsa6/v110/i13/p6158_s1},
	doi = {doi:10.1063/1.478522},
	number = {13},
	urldate = {2013-05-18},
	journal = {J. Chem. Phys.},
	author = {Adamo, Carlo and Barone, Vincenzo},
	month = apr,
	year = {1999},
	pages = {6158--6170},
}

@article{perdew_rationale_1996,
	title = {Rationale for {Mixing} {Exact} {Exchange} with {Density} {Functional} {Approximations}},
	volume = {105},
	issn = {00219606},
	url = {http://jcp.aip.org/resource/1/jcpsa6/v105/i22/p9982_s1},
	doi = {doi:10.1063/1.472933},
	number = {22},
	urldate = {2013-05-20},
	journal = {J. Chem. Phys.},
	author = {Perdew, John P. and Ernzerhof, Matthias and Burke, Kieron},
	month = dec,
	year = {1996},
	pages = {9982--9985},
}

@article{grimme_dispersion-corrected_2016,
	title = {Dispersion-{Corrected} {Mean}-{Field} {Electronic} {Structure} {Methods}},
	volume = {116},
	issn = {0009-2665},
	url = {https://doi.org/10.1021/acs.chemrev.5b00533},
	doi = {10.1021/acs.chemrev.5b00533},
	number = {9},
	urldate = {2025-01-20},
	journal = {Chem. Rev.},
	author = {Grimme, Stefan and Hansen, Andreas and Brandenburg, Jan Gerit and Bannwarth, Christoph},
	month = may,
	year = {2016},
	pages = {5105--5154},
}

@article{hermann_first-principles_2017,
	title = {First-{Principles} {Models} for van der {Waals} {Interactions} in {Molecules} and {Materials}: {Concepts}, {Theory}, and {Applications}},
	volume = {117},
	issn = {0009-2665},
	shorttitle = {First-{Principles} {Models} for van der {Waals} {Interactions} in {Molecules} and {Materials}},
	url = {https://doi.org/10.1021/acs.chemrev.6b00446},
	doi = {10.1021/acs.chemrev.6b00446},
	number = {6},
	urldate = {2018-12-02},
	journal = {Chem. Rev.},
	author = {Hermann, Jan and DiStasio Jr., Robert A. and Tkatchenko, Alexandre},
	month = mar,
	year = {2017},
	pages = {4714--4758},
}

@article{car_unified_1985,
	title = {Unified {Approach} for {Molecular} {Dynamics} and {Density}-{Functional} {Theory}},
	volume = {55},
	url = {http://link.aps.org/doi/10.1103/PhysRevLett.55.2471},
	doi = {10.1103/PhysRevLett.55.2471},
	number = {22},
	urldate = {2012-08-27},
	journal = {Phys. Rev. Lett.},
	author = {Car, R. and Parrinello, M.},
	month = nov,
	year = {1985},
	pages = {2471--2474},
}

@book{marx_ab_2009,
	address = {Cambridge},
	title = {Ab {Initio} {Molecular} {Dynamics}: {Basic} {Theory} and {Advanced} {Methods}},
	shorttitle = {Ab {Initio} {Molecular} {Dynamics}},
	publisher = {Cambridge University Press},
	author = {Marx, Dominik and Hutter, J{\"u}rg},
	month = apr,
	year = {2009},
}

@article{giannozzi_quantum_2020,
	title = {Quantum {ESPRESSO} toward the exascale},
	volume = {152},
	issn = {0021-9606},
	url = {https://aip-scitation-org.proxy.library.cornell.edu/doi/full/10.1063/5.0005082},
	doi = {10.1063/5.0005082},
	number = {15},
	urldate = {2021-09-22},
	journal = {J. Chem. Phys.},
	author = {Giannozzi, Paolo and Baseggio, Oscar and Bonf{\`a}, Pietro and Brunato, Davide and Car, Roberto and Carnimeo, Ivan and Cavazzoni, Carlo and de Gironcoli, Stefano and Delugas, Pietro and Ferrari Ruffino, Fabrizio and Ferretti, Andrea and Marzari, Nicola and Timrov, Iurii and Urru, Andrea and Baroni, Stefano},
	month = apr,
	year = {2020},
	pages = {154105},
}

@article{giannozzi_advanced_2017,
	title = {Advanced capabilities for materials modelling with {Quantum} {ESPRESSO}},
	volume = {29},
	issn = {0953-8984},
	url = {http://stacks.iop.org/0953-8984/29/i=46/a=465901},
	doi = {10.1088/1361-648X/aa8f79},
	language = {en},
	number = {46},
	urldate = {2018-01-06},
	journal = {J. Phys.: Condens. Matter},
	author = {Giannozzi, P. and Andreussi, O. and Brumme, T. and Bunau, O. and Nardelli, M. Buongiorno and Calandra, M. and Car, R. and Cavazzoni, C. and Ceresoli, D. and Cococcioni, M. and Colonna, N. and Carnimeo, I. and Corso, A. Dal and de Gironcoli, S. and Delugas, P. and DiStasio Jr., R. A. and Ferretti, A. and Floris, A. and Fratesi, G. and Fugallo, G. and Gebauer, R. and Gerstmann, U. and Giustino, F. and Gorni, T. and Jia, J. and Kawamura, M. and Ko, H.-Y. and Kokalj, A. and K{\"u}{\c c}{\"u}kbenli, E. and Lazzeri, M. and Marsili, M. and Marzari, N. and Mauri, F. and Nguyen, N. L. and Nguyen, H.-V. and Otero-de-la-Roza, A. and Paulatto, L. and Ponc{\'e}, S. and Rocca, D. and Sabatini, R. and Santra, B. and Schlipf, M. and Seitsonen, A. P. and Smogunov, A. and Timrov, I. and Thonhauser, T. and Umari, P. and Vast, N. and Wu, X. and Baroni, S.},
	year = {2017},
	pages = {465901},
}

@article{giannozzi_quantum_2009,
	title = {{QUANTUM} {ESPRESSO}: a modular and open-source software project for quantum simulations of materials},
	volume = {21},
	issn = {0953-8984, 1361-648X},
	shorttitle = {{QUANTUM} {ESPRESSO}},
	url = {http://iopscience.iop.org/0953-8984/21/39/395502},
	doi = {10.1088/0953-8984/21/39/395502},
	number = {39},
	urldate = {2012-09-13},
	journal = {J. Phys.: Condens. Matter},
	author = {Giannozzi, Paolo and Baroni, Stefano and Bonini, Nicola and Calandra, Matteo and Car, Roberto and Cavazzoni, Carlo and Ceresoli, Davide and Chiarotti, Guido L and Cococcioni, Matteo and Dabo, Ismaila and Dal Corso, Andrea and de Gironcoli, Stefano and Fabris, Stefano and Fratesi, Guido and Gebauer, Ralph and Gerstmann, Uwe and Gougoussis, Christos and Kokalj, Anton and Lazzeri, Michele and Martin-Samos, Layla and Marzari, Nicola and Mauri, Francesco and Mazzarello, Riccardo and Paolini, Stefano and Pasquarello, Alfredo and Paulatto, Lorenzo and Sbraccia, Carlo and Scandolo, Sandro and Sclauzero, Gabriele and Seitsonen, Ari P and Smogunov, Alexander and Umari, Paolo and Wentzcovitch, Renata M},
	month = sep,
	year = {2009},
	pages = {395502},
}

@article{gartner_modeling_2019,
	title = {Modeling and {Simulations} of {Polymers}: {A} {Roadmap}},
	volume = {52},
	issn = {0024-9297},
	shorttitle = {Modeling and {Simulations} of {Polymers}},
	url = {https://doi.org/10.1021/acs.macromol.8b01836},
	doi = {10.1021/acs.macromol.8b01836},
	number = {3},
	urldate = {2025-03-05},
	journal = {Macromolecules},
	author = {Gartner III, Thomas E. and Jayaraman, Arthi},
	month = feb,
	year = {2019},
	pages = {755--786},
}

@article{schmid_understanding_2023,
	title = {Understanding and {Modeling} {Polymers}: {The} {Challenge} of {Multiple} {Scales}},
	volume = {3},
	shorttitle = {Understanding and {Modeling} {Polymers}},
	url = {https://doi.org/10.1021/acspolymersau.2c00049},
	doi = {10.1021/acspolymersau.2c00049},
	number = {1},
	urldate = {2025-03-05},
	journal = {ACS Polym. Au},
	author = {Schmid, Friederike},
	month = feb,
	year = {2023},
	pages = {28--58},
}

@article{olsson_ab_2017,
	title = {\textit{{Ab}} initio and classical atomistic modelling of structure and defects in crystalline orthorhombic polyethylene: {Twin} boundaries, slip interfaces, and nature of barriers},
	volume = {121},
	issn = {0032-3861},
	shorttitle = {\textit{{Ab}} initio and classical atomistic modelling of structure and defects in crystalline orthorhombic polyethylene},
	url = {https://www.sciencedirect.com/science/article/pii/S0032386117305682},
	doi = {10.1016/j.polymer.2017.06.008},
	urldate = {2025-03-05},
	journal = {Polymer},
	author = {Olsson, P{\"a}r A. T. and Schr{\"o}der, Elsebeth and Hyldgaard, Per and Kroon, Martin and Andreasson, Eskil and Bergvall, Erik},
	month = jul,
	year = {2017},
	pages = {234--246},
}

@article{liu_how_2012,
	title = {How {Critical} {Are} the van der {Waals} {Interactions} in {Polymer} {Crystals}?},
	volume = {116},
	issn = {1089-5639},
	url = {https://doi.org/10.1021/jp3005844},
	doi = {10.1021/jp3005844},
	number = {37},
	urldate = {2025-03-06},
	journal = {J. Phys. Chem. A},
	author = {Liu, Chun-Sheng and Pilania, Ghanshyam and Wang, Chenchen and Ramprasad, Ramamurthy},
	month = sep,
	year = {2012},
	pages = {9347--9352},
}

@article{perdew_generalized_1996,
	title = {Generalized {Gradient} {Approximation} {Made} {Simple}},
	volume = {77},
	url = {http://link.aps.org/doi/10.1103/PhysRevLett.77.3865},
	doi = {10.1103/PhysRevLett.77.3865},
	number = {18},
	urldate = {2013-05-18},
	journal = {Phys. Rev. Lett.},
	author = {Perdew, John P. and Burke, Kieron and Ernzerhof, Matthias},
	month = oct,
	year = {1996},
	pages = {3865--3868},
}

@inproceedings{perdew_jacobs_2001,
	address = {Melville},
	title = {Jacob{\textquoteright}s {Ladder} of {Density} {Functional} {Approximations} for the {Exchange}-{Correlation} {Energy}},
	volume = {577},
	url = {http://scitation.aip.org/content/aip/proceeding/aipcp/10.1063/1.1390175},
	urldate = {2015-03-26},
	booktitle = {Density {Functional} {Theory} and {Its} {Application} to {Materials}: {Antwerp}, {Belgium}, {Jun}. 8-10, 2000},
	publisher = {AIP Publishing},
	author = {Perdew, John P. and Schmidt, Karla},
	editor = {Van Doren, Victor E. and Van Alsenoy, C. and Geerlings, P.},
	year = {2001},
	pages = {1--20},
}

@article{huan_polymer_2020,
	title = {Polymer {Structure} {Prediction} from {First} {Principles}},
	volume = {11},
	url = {https://doi.org/10.1021/acs.jpclett.0c01553},
	doi = {10.1021/acs.jpclett.0c01553},
	number = {15},
	urldate = {2022-08-30},
	journal = {J. Phys. Chem. Lett.},
	author = {Huan, Tran Doan and Ramprasad, Rampi},
	month = aug,
	year = {2020},
	pages = {5823--5829},
}

@article{aggarwal_polyethylene_1957,
	title = {Polyethylene: {Preparation}, {Structure}, {And} {Properties}},
	volume = {57},
	issn = {0009-2665, 1520-6890},
	shorttitle = {Polyethylene},
	url = {https://pubs.acs.org/doi/abs/10.1021/cr50016a004},
	doi = {10.1021/cr50016a004},
	language = {en},
	number = {4},
	urldate = {2024-12-02},
	journal = {Chem. Rev.},
	author = {Aggarwal, Sundar L. and Sweeting, Orville J.},
	month = aug,
	year = {1957},
	pages = {665--742},
}

@article{bussi_canonical_2007,
	title = {Canonical sampling through velocity rescaling},
	volume = {126},
	issn = {0021-9606, 1089-7690},
	url = {http://scitation.aip.org/content/aip/journal/jcp/126/1/10.1063/1.2408420},
	doi = {10.1063/1.2408420},
	number = {1},
	urldate = {2015-10-28},
	journal = {J. Chem. Phys.},
	author = {Bussi, Giovanni and Donadio, Davide and Parrinello, Michele},
	month = jan,
	year = {2007},
	pages = {014101},
}

@article{hong_first-principles-based_2021,
	title = {First-{Principles}-{Based} {Machine}-{Learning} {Molecular} {Dynamics} for {Crystalline} {Polymers} with van der {Waals} {Interactions}},
	volume = {12},
	url = {https://doi.org/10.1021/acs.jpclett.1c01140},
	doi = {10.1021/acs.jpclett.1c01140},
	number = {25},
	urldate = {2024-05-19},
	journal = {J. Phys. Chem. Lett.},
	author = {Hong, Sung Jun and Chun, Hoje and Lee, Jehyun and Kim, Byung-Hyun and Seo, Min Ho and Kang, Joonhee and Han, Byungchan},
	month = jul,
	year = {2021},
	pages = {6000--6006},
}

@article{plimpton_fast_1995,
	title = {Fast {Parallel} {Algorithms} for {Short}-{Range} {Molecular} {Dynamics}},
	volume = {117},
	issn = {0021-9991},
	url = {http://www.sciencedirect.com/science/article/pii/S002199918571039X},
	doi = {10.1006/jcph.1995.1039},
	number = {1},
	urldate = {2019-04-05},
	journal = {J. Comput. Phys.},
	author = {Plimpton, Steve},
	month = mar,
	year = {1995},
	pages = {1--19},
}

@article{thompson_lammps_2022,
	title = {{LAMMPS} - a flexible simulation tool for particle-based materials modeling at the atomic, meso, and continuum scales},
	volume = {271},
	issn = {0010-4655},
	url = {https://www.sciencedirect.com/science/article/pii/S0010465521002836},
	doi = {10.1016/j.cpc.2021.108171},
	urldate = {2022-02-10},
	journal = {Comput. Phys. Commun.},
	author = {Thompson, Aidan P. and Aktulga, H. Metin and Berger, Richard and Bolintineanu, Dan S. and Brown, W. Michael and Crozier, Paul S. and in 't Veld, Pieter J. and Kohlmeyer, Axel and Moore, Stan G. and Nguyen, Trung Dac and Shan, Ray and Stevens, Mark J. and Tranchida, Julien and Trott, Christian and Plimpton, Steven J.},
	month = feb,
	year = {2022},
	pages = {108171},
}

@article{martinez_packmol_2009,
	title = {{PACKMOL}: {A} package for building initial configurations for molecular dynamics simulations},
	volume = {30},
	issn = {1096-987X},
	shorttitle = {{PACKMOL}},
	url = {http://onlinelibrary.wiley.com/doi/abs/10.1002/jcc.21224},
	doi = {10.1002/jcc.21224},
	language = {de},
	number = {13},
	urldate = {2022-10-13},
	journal = {J. Comput. Chem.},
	author = {Mart{\'i}nez, L. and Andrade, R. and Birgin, E. G. and Mart{\'i}nez, J. M.},
	year = {2009},
	pages = {2157--2164},
}

@article{unke_machine_2021,
	title = {Machine {Learning} {Force} {Fields}},
	volume = {121},
	issn = {0009-2665},
	url = {https://doi.org/10.1021/acs.chemrev.0c01111},
	doi = {10.1021/acs.chemrev.0c01111},
	number = {16},
	urldate = {2022-05-28},
	journal = {Chem. Rev.},
	author = {Unke, Oliver T. and Chmiela, Stefan and Sauceda, Huziel E. and Gastegger, Michael and Poltavsky, Igor and Sch{\"u}tt, Kristof T. and Tkatchenko, Alexandre and M{\"u}ller, Klaus-Robert},
	month = aug,
	year = {2021},
	pages = {10142--10186},
}

@article{meuwly_machine_2021,
	title = {Machine {Learning} for {Chemical} {Reactions}},
	volume = {121},
	issn = {0009-2665},
	url = {https://doi.org/10.1021/acs.chemrev.1c00033},
	doi = {10.1021/acs.chemrev.1c00033},
	number = {16},
	urldate = {2022-07-10},
	journal = {Chem. Rev.},
	author = {Meuwly, Markus},
	month = aug,
	year = {2021},
	pages = {10218--10239},
}

@article{huang_ab_2021,
	title = {Ab {Initio} {Machine} {Learning} in {Chemical} {Compound} {Space}},
	volume = {121},
	issn = {0009-2665},
	url = {https://doi.org/10.1021/acs.chemrev.0c01303},
	doi = {10.1021/acs.chemrev.0c01303},
	number = {16},
	urldate = {2022-07-10},
	journal = {Chem. Rev.},
	author = {Huang, Bing and von Lilienfeld, O. Anatole},
	month = aug,
	year = {2021},
	pages = {10001--10036},
}

@article{wen_deep_2022,
	title = {Deep potentials for materials science},
	volume = {1},
	issn = {2752-5724},
	url = {https://dx.doi.org/10.1088/2752-5724/ac681d},
	doi = {10.1088/2752-5724/ac681d},
	number = {2},
	urldate = {2024-11-11},
	journal = {Mater. Futures},
	author = {Wen, Tongqi and Zhang, Linfeng and Wang, Han and E, Weinan and Srolovitz, David J.},
	month = may,
	year = {2022},
	pages = {022601},
}

@article{zhang_phase_2021,
	title = {Phase {Diagram} of a {Deep} {Potential} {Water} {Model}},
	volume = {126},
	url = {https://link.aps.org/doi/10.1103/PhysRevLett.126.236001},
	doi = {10.1103/PhysRevLett.126.236001},
	number = {23},
	urldate = {2021-10-20},
	journal = {Phys. Rev. Lett.},
	author = {Zhang, Linfeng and Wang, Han and Car, Roberto and E, Weinan},
	month = jun,
	year = {2021},
	pages = {236001},
}

@article{damle_compressed_2015,
	title = {Compressed {Representation} of {Kohn}{\textendash}{Sham} {Orbitals} via {Selected} {Columns} of the {Density} {Matrix}},
	volume = {11},
	issn = {1549-9618},
	url = {https://doi.org/10.1021/ct500985f},
	doi = {10.1021/ct500985f},
	number = {4},
	urldate = {2018-05-01},
	journal = {J. Chem. Theory Comput.},
	author = {Damle, Anil and Lin, Lin and Ying, Lexing},
	month = apr,
	year = {2015},
	pages = {1463--1469},
}

@article{kohn_self-consistent_1965,
	title = {Self-{Consistent} {Equations} {Including} {Exchange} and {Correlation} {Effects}},
	volume = {140},
	url = {http://link.aps.org/doi/10.1103/PhysRev.140.A1133},
	doi = {10.1103/PhysRev.140.A1133},
	number = {4A},
	urldate = {2013-03-26},
	journal = {Phys. Rev.},
	author = {Kohn, W. and Sham, L. J.},
	month = nov,
	year = {1965},
	pages = {A1133--A1138},
}

@article{hohenberg_inhomogeneous_1964,
	title = {Inhomogeneous {Electron} {Gas}},
	volume = {136},
	url = {http://link.aps.org/doi/10.1103/PhysRev.136.B864},
	doi = {10.1103/PhysRev.136.B864},
	number = {3B},
	urldate = {2013-03-26},
	journal = {Phys. Rev.},
	author = {Hohenberg, P. and Kohn, W.},
	month = nov,
	year = {1964},
	pages = {B864--B871},
}

@article{jones_density_1989,
	title = {The {Density} {Functional} {Formalism}, {Its} {Applications} and {Prospects}},
	volume = {61},
	url = {http://link.aps.org/doi/10.1103/RevModPhys.61.689},
	doi = {10.1103/RevModPhys.61.689},
	number = {3},
	urldate = {2015-03-26},
	journal = {Rev. Mod. Phys.},
	author = {Jones, R. O. and Gunnarsson, O.},
	month = jul,
	year = {1989},
	pages = {689--746},
}

@book{parr_density-functional_1989,
	address = {New York},
	title = {Density-{Functional} {Theory} of {Atoms} and {Molecules}},
	isbn = {978-0-19-535773-8},
	publisher = {Oxford University Press},
	author = {Parr, Robert G. and Yang, Weitao},
	month = apr,
	year = {1989},
}

@article{gygi_self-consistent_1986,
	title = {Self-consistent {Hartree}-{Fock} and screened-exchange calculations in solids: {Application} to silicon},
	volume = {34},
	shorttitle = {Self-consistent {Hartree}-{Fock} and screened-exchange calculations in solids},
	url = {http://link.aps.org/doi/10.1103/PhysRevB.34.4405},
	doi = {10.1103/PhysRevB.34.4405},
	number = {6},
	urldate = {2014-09-02},
	journal = {Phys. Rev. B},
	author = {Gygi, F. and Baldereschi, A.},
	month = sep,
	year = {1986},
	pages = {4405--4408},
}

@article{lin_adaptively_2016,
	title = {Adaptively {Compressed} {Exchange} {Operator}},
	volume = {12},
	issn = {1549-9618},
	url = {http://dx.doi.org/10.1021/acs.jctc.6b00092},
	doi = {10.1021/acs.jctc.6b00092},
	number = {5},
	urldate = {2017-05-22},
	journal = {J. Chem. Theory Comput.},
	author = {Lin, Lin},
	month = may,
	year = {2016},
	pages = {2242--2249},
}

@article{lee_development_1988,
	title = {Development of the {Colle}-{Salvetti} correlation-energy formula into a functional of the electron density},
	volume = {37},
	url = {http://journals.aps.org/prb/abstract/10.1103/PhysRevB.37.785},
	number = {2},
	urldate = {2015-04-28},
	journal = {Phys. Rev. B},
	author = {Lee, Chengteh and Yang, Weitao and Parr, Robert G.},
	year = {1988},
	pages = {785--789},
}

@article{boero_first_2000,
	title = {First {Principles} {Study} of {Propene} {Polymerization} in {Ziegler}-{Natta} {Heterogeneous} {Catalysis}},
	volume = {122},
	issn = {0002-7863},
	url = {https://doi.org/10.1021/ja990913x},
	doi = {10.1021/ja990913x},
	number = {3},
	urldate = {2022-10-11},
	journal = {J. Am. Chem. Soc.},
	author = {Boero, Mauro and Parrinello, Michele and H{\"u}ffer, Stephan and Weiss, Horst},
	month = jan,
	year = {2000},
	pages = {501--509},
}

@article{ferretti_ab_2004,
	title = {Ab initio study of transport parameters in polymer crystals},
	volume = {69},
	url = {https://link.aps.org/doi/10.1103/PhysRevB.69.205205},
	doi = {10.1103/PhysRevB.69.205205},
	number = {20},
	urldate = {2025-07-30},
	journal = {Phys. Rev. B},
	author = {Ferretti, Andrea and Ruini, Alice and Bussi, Giovanni and Molinari, Elisa and Caldas, Marilia J.},
	month = may,
	year = {2004},
	pages = {205205},
}

@article{xue_ab_2019,
	title = {Ab \textit{initio} calculations for crystalline {PEO6}:{LiPF6} polymer electrolytes},
	volume = {160},
	issn = {0927-0256},
	shorttitle = {Ab \textit{initio} calculations for crystalline {PEO6}},
	url = {https://www.sciencedirect.com/science/article/pii/S0927025619300072},
	doi = {10.1016/j.commatsci.2019.01.007},
	urldate = {2025-07-30},
	journal = {Comput. Mater. Sci.},
	author = {Xue, Sha and Teeters, Dale and Crunkleton, Daniel W. and Wang, Sanwu},
	month = apr,
	year = {2019},
	pages = {173--179},
}

@article{kurita_crystalline_2018,
	title = {Crystalline {Moduli} of {Polymers}, {Evaluated} from {Density} {Functional} {Theory} {Calculations} under {Periodic} {Boundary} {Conditions}},
	volume = {3},
	url = {https://doi.org/10.1021/acsomega.8b00506},
	doi = {10.1021/acsomega.8b00506},
	number = {5},
	urldate = {2025-07-30},
	journal = {ACS Omega},
	author = {Kurita, Taiga and Fukuda, Yuichiro and Takahashi, Morihiro and Sasanuma, Yuji},
	month = may,
	year = {2018},
	pages = {4824--4835},
}

@article{fontana_high-pressure_2007,
	title = {High-pressure crystalline polyethylene studied by x-ray diffraction and ab initio simulations},
	volume = {75},
	url = {https://link.aps.org/doi/10.1103/PhysRevB.75.174112},
	doi = {10.1103/PhysRevB.75.174112},
	number = {17},
	urldate = {2025-07-30},
	journal = {Phys. Rev. B},
	author = {Fontana, L. and Vinh, Diep Q. and Santoro, M. and Scandolo, S. and Gorelli, F. A. and Bini, R. and Hanfland, M.},
	month = may,
	year = {2007},
	pages = {174112},
}

@article{beyer_mechanochemistry_2005,
	title = {Mechanochemistry: {The} {Mechanical} {Activation} of {Covalent} {Bonds}},
	volume = {105},
	issn = {0009-2665},
	shorttitle = {Mechanochemistry},
	url = {https://doi.org/10.1021/cr030697h},
	doi = {10.1021/cr030697h},
	number = {8},
	urldate = {2025-08-04},
	journal = {Chem. Rev.},
	author = {Beyer, Martin K. and Clausen-Schaumann, Hauke},
	month = aug,
	year = {2005},
	pages = {2921--2948},
}

@article{ribas-arino_covalent_2012,
	title = {Covalent {Mechanochemistry}: {Theoretical} {Concepts} and {Computational} {Tools} with {Applications} to {Molecular} {Nanomechanics}},
	volume = {112},
	issn = {0009-2665},
	shorttitle = {Covalent {Mechanochemistry}},
	url = {https://doi.org/10.1021/cr200399q},
	doi = {10.1021/cr200399q},
	number = {10},
	urldate = {2025-08-04},
	journal = {Chem. Rev.},
	author = {Ribas-Arino, Jordi and Marx, Dominik},
	month = oct,
	year = {2012},
	pages = {5412--5487},
}

@article{vaschetto_first-principles_1999,
	title = {First-{Principles} {Calculations} of {Pyridines}: {From} {Monomer} to {Polymer}},
	volume = {103},
	issn = {1089-5639},
	shorttitle = {First-{Principles} {Calculations} of {Pyridines}},
	url = {https://doi.org/10.1021/jp992204b},
	doi = {10.1021/jp992204b},
	number = {50},
	urldate = {2025-08-06},
	journal = {J. Phys. Chem. A},
	author = {Vaschetto, Mariana E. and Retamal, Bernardo A. and Monkman, Andrew P. and Springborg, Michael},
	month = dec,
	year = {1999},
	pages = {11096--11103},
}

@article{salzner_accurate_1998,
	title = {Accurate {Method} for {Obtaining} {Band} {Gaps} in {Conducting} {Polymers} {Using} a {DFT}/{Hybrid} {Approach}},
	volume = {102},
	issn = {1089-5639},
	url = {https://doi.org/10.1021/jp971652l},
	doi = {10.1021/jp971652l},
	number = {15},
	urldate = {2025-08-06},
	journal = {J. Phys. Chem. A},
	author = {Salzner, U. and Pickup, P. G. and Poirier, R. A. and Lagowski, J. B.},
	month = apr,
	year = {1998},
	pages = {2572--2578},
}

@article{salzner_design_1997,
	title = {Design of low band gap polymers employing density functional theory{\textemdash}hybrid functionals ameliorate band gap problem},
	volume = {18},
	copyright = {Copyright {\textcopyright} 1997 John Wiley \& Sons, Inc.},
	issn = {1096-987X},
	url = {https://onlinelibrary.wiley.com/doi/abs/10.1002/%28SICI%291096-987X%2819971130%2918%3A15%3C1943%3A%3AAID-JCC9%3E3.0.CO%3B2-O},
	doi = {10.1002/(SICI)1096-987X(19971130)18:15<1943::AID-JCC9>3.0.CO;2-O},
	number = {15},
	urldate = {2025-08-06},
	journal = {J. Comput. Chem.},
	author = {Salzner, U. and Lagowski, J. B. and Pickup, P. G. and Poirier, R. A.},
	year = {1997},
	pages = {1943--1953},
}

@article{salzner_comparison_1998,
	title = {Comparison of geometries and electronic structures of polyacetylene, polyborole, polycyclopentadiene, polypyrrole, polyfuran, polysilole, polyphosphole, polythiophene, polyselenophene and polytellurophene},
	volume = {96},
	issn = {0379-6779},
	url = {https://www.sciencedirect.com/science/article/pii/S0379677998000848},
	doi = {10.1016/S0379-6779(98)00084-8},
	number = {3},
	urldate = {2025-08-06},
	journal = {Synth. Met.},
	author = {Salzner, U. and Lagowski, J. B. and Pickup, P. G. and Poirier, R. A.},
	month = aug,
	year = {1998},
	pages = {177--189},
}

@article{de_oliveira_energy_2000,
	title = {Energy {Gaps} of a,a'-{Substituted} {Oligothiophenes} from {Semiempirical}, {Ab} {Initio}, and {Density} {Functional} {Methods}},
	volume = {104},
	issn = {1089-5639},
	url = {https://doi.org/10.1021/jp001252p},
	doi = {10.1021/jp001252p},
	number = {35},
	urldate = {2025-08-06},
	journal = {J. Phys. Chem. A},
	author = {De Oliveira, Marcos A. and Duarte, H{\'e}lio A. and Pernaut, Jean-Michel and De Almeida, Wagner B.},
	month = sep,
	year = {2000},
	pages = {8256--8262},
}

@article{zade_short_2011,
	title = {From {Short} {Conjugated} {Oligomers} to {Conjugated} {Polymers}. {Lessons} from {Studies} on {Long} {Conjugated} {Oligomers}},
	volume = {44},
	issn = {0001-4842},
	url = {https://doi.org/10.1021/ar1000555},
	doi = {10.1021/ar1000555},
	number = {1},
	urldate = {2025-08-07},
	journal = {Acc. Chem. Res.},
	author = {Zade, Sanjio S. and Zamoshchik, Natalia and Bendikov, Michael},
	month = jan,
	year = {2011},
	pages = {14--24},
}

@article{korzdorfer_organic_2014,
	title = {Organic {Electronic} {Materials}: {Recent} {Advances} in the {DFT} {Description} of the {Ground} and {Excited} {States} {Using} {Tuned} {Range}-{Separated} {Hybrid} {Functionals}},
	volume = {47},
	issn = {0001-4842},
	shorttitle = {Organic {Electronic} {Materials}},
	url = {https://doi.org/10.1021/ar500021t},
	doi = {10.1021/ar500021t},
	number = {11},
	urldate = {2025-08-07},
	journal = {Acc. Chem. Res.},
	author = {K{\"o}rzd{\"o}rfer, Thomas and Br{\'e}das, Jean-Luc},
	month = nov,
	year = {2014},
	pages = {3284--3291},
}

@article{bernasconi_solid-state_1997,
	title = {Solid-{State} {Polymerization} of {Acetylene} under {Pressure}: {Ab} {Initio} {Simulation}},
	volume = {78},
	shorttitle = {Solid-{State} {Polymerization} of {Acetylene} under {Pressure}},
	url = {https://link.aps.org/doi/10.1103/PhysRevLett.78.2008},
	doi = {10.1103/PhysRevLett.78.2008},
	number = {10},
	urldate = {2022-03-15},
	journal = {Phys. Rev. Lett.},
	author = {Bernasconi, M. and Chiarotti, G. L. and Focher, P. and Parrinello, M. and Tosatti, E.},
	month = mar,
	year = {1997},
	pages = {2008--2011},
}

@article{grimme_consistent_2010,
	title = {A consistent and accurate ab initio parametrization of density functional dispersion correction ({DFT}-{D}) for the 94 elements {H}-{Pu}},
	volume = {132},
	issn = {0021-9606},
	url = {https://doi.org/10.1063/1.3382344},
	doi = {10.1063/1.3382344},
	number = {15},
	urldate = {2025-10-13},
	journal = {J. Chem. Phys.},
	author = {Grimme, Stefan and Antony, Jens and Ehrlich, Stephan and Krieg, Helge},
	month = apr,
	year = {2010},
	pages = {154104},
}

@book{frenkel_understanding_2001,
	address = {New York},
	title = {Understanding molecular simulation: from algorithms to applications},
	volume = {1},
	shorttitle = {Understanding molecular simulation},
	url = {https://books.google.com/books?hl=en&lr=&id=5qTzldS9ROIC&oi=fnd&pg=PP2&dq=frenkel+smit&ots=nFQHVo38Xi&sig=aRD7WP1NtbFlqq5huDFBD3rnjT0},
	urldate = {2016-04-08},
	publisher = {Academic press},
	author = {Frenkel, Daan and Smit, Berend},
	year = {2001},
}

@book{allen_computer_1989,
	title = {Computer simulation of liquids},
	urldate = {2016-03-28},
	publisher = {Oxford university press},
	author = {Allen, Mike P. and Tildesley, Dominic J.},
	year = {1989},
}

@article{van_der_giessen_roadmap_2020,
	title = {Roadmap on multiscale materials modeling},
	volume = {28},
	issn = {0965-0393},
	url = {https://doi.org/10.1088/1361-651X/ab7150},
	doi = {10.1088/1361-651X/ab7150},
	language = {en},
	number = {4},
	urldate = {2026-02-05},
	journal = {Modelling Simul. Mater. Sci. Eng.},
	author = {van der Giessen, Erik and Schultz, Peter A and Bertin, Nicolas and Bulatov, Vasily V and Cai, Wei and Cs{\'a}nyi, G{\'a}bor and Foiles, Stephen M and Geers, M G D and Gonz{\'a}lez, Carlos and H{\"u}tter, Markus and Kim, Woo Kyun and Kochmann, Dennis M and LLorca, Javier and Mattsson, Ann E and Rottler, J{\"o}rg and Shluger, Alexander and Sills, Ryan B and Steinbach, Ingo and Strachan, Alejandro and Tadmor, Ellad B},
	month = mar,
	year = {2020},
	pages = {043001},
}

@article{zeng_multiscale_2008,
	title = {Multiscale modeling and simulation of polymer nanocomposites},
	volume = {33},
	issn = {0079-6700},
	url = {https://www.sciencedirect.com/science/article/pii/S0079670007001049},
	doi = {10.1016/j.progpolymsci.2007.09.002},
	number = {2},
	urldate = {2026-02-05},
	journal = {Progress in Polymer Science},
	author = {Zeng, Q. H. and Yu, A. B. and Lu, G. Q.},
	month = feb,
	year = {2008},
	pages = {191--269},
}

@book{engquist_multiscale_2009,
	title = {Multiscale {Modeling} and {Simulation} in {Science}},
	isbn = {978-3-540-88857-4},
	publisher = {Springer Science \& Business Media},
	author = {Engquist, Bj{\"o}rn and L{\"o}tstedt, Per and Runborg, Olof},
	month = feb,
	year = {2009},
}

@article{becke_density-functional_1988,
	title = {Density-{Functional} {Exchange}-{Energy} {Approximation} with {Correct} {Asymptotic} {Behavior}},
	volume = {38},
	url = {http://link.aps.org/doi/10.1103/PhysRevA.38.3098},
	doi = {10.1103/PhysRevA.38.3098},
	number = {6},
	urldate = {2014-04-04},
	journal = {Phys. Rev. A},
	author = {Becke, A. D.},
	month = sep,
	year = {1988},
	pages = {3098--3100},
}

@article{perdew_comparison_1996,
	title = {Comparison shopping for a gradient-corrected density functional},
	volume = {57},
	issn = {1097-461X},
	url = {https://onlinelibrary.wiley.com/doi/abs/10.1002/%28SICI%291097-461X%281996%2957%3A3%3C309%3A%3AAID-QUA4%3E3.0.CO%3B2-1},
	doi = {10.1002/(SICI)1097-461X(1996)57:3<309::AID-QUA4>3.0.CO;2-1},
	number = {3},
	urldate = {2026-02-16},
	journal = {Int. J. Quantum Chem.},
	author = {Perdew, John P. and Burke, Kieron},
	year = {1996},
	pages = {309--319},
}

@article{hamann_optimized_2013,
	title = {Optimized norm-conserving {Vanderbilt} pseudopotentials},
	volume = {88},
	url = {https://link.aps.org/doi/10.1103/PhysRevB.88.085117},
	doi = {10.1103/PhysRevB.88.085117},
	number = {8},
	urldate = {2019-12-19},
	journal = {Phys. Rev. B},
	author = {Hamann, D. R.},
	month = aug,
	year = {2013},
	pages = {085117},
}

@article{schlipf_optimization_2015,
	title = {Optimization algorithm for the generation of {ONCV} pseudopotentials},
	volume = {196},
	issn = {0010-4655},
	url = {http://www.sciencedirect.com/science/article/pii/S0010465515001897},
	doi = {10.1016/j.cpc.2015.05.011},
	urldate = {2019-09-13},
	journal = {Comput. Phys. Commun.},
	author = {Schlipf, Martin and Gygi, Fran{\c c}ois},
	month = nov,
	year = {2015},
	pages = {36--44},
}

@article{deringer_machine_2019,
	title = {Machine {Learning} {Interatomic} {Potentials} as {Emerging} {Tools} for {Materials} {Science}},
	volume = {31},
	copyright = {{\textcopyright} 2019 WILEY-VCH Verlag GmbH \& Co. KGaA, Weinheim},
	issn = {1521-4095},
	url = {https://onlinelibrary.wiley.com/doi/abs/10.1002/adma.201902765},
	doi = {10.1002/adma.201902765},
	language = {en},
	number = {46},
	urldate = {2026-07-02},
	journal = {Adv. Mater.},
	author = {Deringer, Volker L. and Caro, Miguel A. and Cs{\'a}nyi, G{\'a}bor},
	year = {2019},
	pages = {1902765},
}

@article{behler_four_2021,
	title = {Four {Generations} of {High}-{Dimensional} {Neural} {Network} {Potentials}},
	volume = {121},
	issn = {0009-2665},
	url = {https://doi.org/10.1021/acs.chemrev.0c00868},
	doi = {10.1021/acs.chemrev.0c00868},
	number = {16},
	urldate = {2026-07-02},
	journal = {Chem. Rev.},
	publisher = {American Chemical Society},
	author = {Behler, J{\"o}rg},
	month = aug,
	year = {2021},
	pages = {10037--10072},
}

@article{ceriotti_introduction_2021,
	title = {Introduction: {Machine} {Learning} at the {Atomic} {Scale}},
	volume = {121},
	issn = {0009-2665},
	shorttitle = {Introduction},
	url = {https://doi.org/10.1021/acs.chemrev.1c00598},
	doi = {10.1021/acs.chemrev.1c00598},
	number = {16},
	urldate = {2026-07-02},
	journal = {Chem. Rev.},
	publisher = {American Chemical Society},
	author = {Ceriotti, Michele and Clementi, Cecilia and Anatole von Lilienfeld, O.},
	month = aug,
	year = {2021},
	pages = {9719--9721},
}

@article{ko_general-purpose_2021,
	title = {General-{Purpose} {Machine} {Learning} {Potentials} {Capturing} {Nonlocal} {Charge} {Transfer}},
	volume = {54},
	issn = {0001-4842},
	url = {https://doi.org/10.1021/acs.accounts.0c00689},
	doi = {10.1021/acs.accounts.0c00689},
	number = {4},
	urldate = {2026-07-02},
	journal = {Acc. Chem. Res.},
	publisher = {American Chemical Society},
	author = {Ko, Tsz Wai and Finkler, Jonas A. and Goedecker, Stefan and Behler, J{\"o}rg},
	month = feb,
	year = {2021},
	pages = {808--817},
}

@article{kocer_neural_2022,
	title = {Neural {Network} {Potentials}: {A} {Concise} {Overview} of {Methods}},
	volume = {73},
	issn = {0066-426X, 1545-1593},
	shorttitle = {Neural {Network} {Potentials}},
	url = {https://www.annualreviews.org/content/journals/10.1146/annurev-physchem-082720-034254},
	doi = {10.1146/annurev-physchem-082720-034254},
	language = {en},
	number = {Volume 73, 2022},
	urldate = {2026-07-02},
	journal = {Annu. Rev. Phys. Chem.},
	publisher = {Annual Reviews},
	author = {Kocer, Emir and Ko, Tsz Wai and Behler, J{\"o}rg},
	month = apr,
	year = {2022},
	pages = {163--186},
}

@misc{simm_simpoly_2025,
	title = {{SimPoly}: {Simulation} of {Polymers} with {Machine} {Learning} {Force} {Fields} {Derived} from {First} {Principles}},
	shorttitle = {{SimPoly}},
	url = {http://arxiv.org/abs/2510.13696},
	doi = {10.48550/arXiv.2510.13696},
	urldate = {2026-07-02},
	publisher = {arXiv},
	author = {Simm, Gregor N. C. and H{\'e}lie, Jean and Schulz, Hannes and Chen, Yicheng and Simeon, Guillem and Kuzina, Anna and Martinez-Baez, Ernesto and Gasparotto, Piero and Tocci, Gabriele and Chen, Chi and Li, Yatao and Cheng, Lixue and Wang, Zun and Nguyen, Bichlien H. and Smith, Jake A. and Sun, Lixin},
	month = oct,
	year = {2025},
	note = {arXiv:2510.13696 [physics]},
}

@article{mieda_comparison_2025,
	title = {Comparison of {Computational} {Methods} for {Simulating} {Depolymerization} {Reaction}},
	volume = {10},
	issn = {2470-1343},
	url = {https://doi.org/10.1021/acsomega.4c09953},
	doi = {10.1021/acsomega.4c09953},
	number = {6},
	urldate = {2026-08-09},
	journal = {ACS Omega},
	author = {Mieda, Shunsuke},
	month = feb,
	year = {2025},
	pages = {5973--5980},
}

@article{ma_understanding_2023,
	title = {Understanding {Rapid} {PET} {Degradation} via {Reactive} {Molecular} {Dynamics} {Simulation} and {Kinetic} {Modeling}},
	volume = {127},
	issn = {1089-5639},
	url = {https://doi.org/10.1021/acs.jpca.3c03717},
	doi = {10.1021/acs.jpca.3c03717},
	number = {35},
	urldate = {2026-08-09},
	journal = {J. Phys. Chem. A},
	author = {Ma, Shuangxiu Max and Zou, Changlong and Chen, Ting-Yeh and Paulson, Joel A. and Lin, Li-Chiang and Bakshi, Bhavik R.},
	month = aug,
	year = {2023},
	pages = {7323--7334},
}

@article{li_modeling-driven_2024,
	title = {Modeling-driven materials by design for conjugated polymers: insights into optoelectronic, conformational, and thermomechanical properties},
	volume = {60},
	issn = {1359-7345},
	shorttitle = {Modeling-driven materials by design for conjugated polymers},
	url = {https://doi.org/10.1039/d4cc03217a},
	doi = {10.1039/d4cc03217a},
	number = {82},
	urldate = {2026-08-09},
	journal = {Chem. Commun.},
	author = {Li, Zhaofan and Tolba, Sara A. and Wang, Yang and Alesadi, Amirhadi and Xia, Wenjie},
	month = oct,
	year = {2024},
	pages = {11625--11641},
}

@article{ma_perspective_2021,
	title = {A {Perspective} on the {Design} of {Ion}-{Containing} {Polymers} for {Polymer} {Electrolyte} {Applications}},
	volume = {125},
	issn = {1520-6106},
	url = {https://doi.org/10.1021/acs.jpcb.0c08707},
	doi = {10.1021/acs.jpcb.0c08707},
	number = {12},
	urldate = {2026-08-09},
	journal = {J. Phys. Chem. B},
	author = {Ma, Boran and Olvera de la Cruz, Monica},
	month = feb,
	year = {2021},
	pages = {3015--3022},
}

@article{tan_bridging_2023,
	title = {Bridging the {Monomer} to {Polymer} {Gap} in {Radical} {Polymer} {Design}},
	volume = {12},
	issn = {2161-1653},
	url = {https://doi.org/10.1021/acsmacrolett.3c00105},
	doi = {10.1021/acsmacrolett.3c00105},
	number = {6},
	urldate = {2026-08-09},
	journal = {ACS Macro Lett.},
	author = {Tan, Ying and Boudouris, Bryan W. and Savoie, Brett M.},
	month = may,
	year = {2023},
	pages = {801--807},
}

@article{qin_organic_2025,
	title = {Organic {Mixed} {Ionic}{\textendash}{Electronic} {Conductors} for {Organic} {Electrochemical} {Transistors}: {Sidechain} {Structure} {Influences} {Ion} {Uptake} and {Functional} {Performance}},
	volume = {26},
	copyright = {{\textcopyright} 2025 The Author(s). ChemPhysChem published by Wiley-VCH GmbH},
	issn = {1439-7641},
	shorttitle = {Organic {Mixed} {Ionic}{\textendash}{Electronic} {Conductors} for {Organic} {Electrochemical} {Transistors}},
	url = {https://onlinelibrary.wiley.com/doi/abs/10.1002/cphc.202500403},
	doi = {10.1002/cphc.202500403},
	language = {en},
	number = {21},
	urldate = {2026-08-09},
	journal = {ChemPhysChem},
	author = {Qin, Siyu and Sun, Zeyuan and Li, Haoxuan and Rahman, Charleen and Gartner III, Thomas E. and Reichmanis, Elsa},
	year = {2025},
	pages = {e202500403},
}

@article{chen_phyneo_2023,
	title = {{PhyNEO}: {A} {Neural}-{Network}-{Enhanced} {Physics}-{Driven} {Force} {Field} {Development} {Workflow} for {Bulk} {Organic} {Molecule} and {Polymer} {Simulations}},
	volume = {20},
	issn = {1549-9618},
	shorttitle = {{PhyNEO}},
	url = {https://doi.org/10.1021/acs.jctc.3c01045},
	doi = {10.1021/acs.jctc.3c01045},
	number = {1},
	urldate = {2026-08-10},
	journal = {J. Chem. Theory Comput.},
	author = {Chen, Junmin and Yu, Kuang},
	month = dec,
	year = {2023},
	pages = {253--265},
}

@article{mohanty_development_2023,
	title = {Development of scalable and generalizable machine learned force field for polymers},
	volume = {13},
	copyright = {2023 The Author(s)},
	issn = {2045-2322},
	url = {https://www.nature.com/articles/s41598-023-43804-5},
	doi = {10.1038/s41598-023-43804-5},
	language = {en},
	number = {1},
	urldate = {2026-08-10},
	journal = {Sci. Rep.},
	publisher = {Nature Publishing Group},
	author = {Mohanty, Shaswat and Stevenson, James and Browning, Andrea R. and Jacobson, Leif and Leswing, Karl and Halls, Mathew D. and Afzal, Mohammad Atif Faiz},
	month = oct,
	year = {2023},
	pages = {17251},
}

@article{hooven_how_2026,
	title = {How {Long} {Is} {Long} {Enough}? {Extrapolation} of {Machine}-{Learning} {Interatomic} {Potentials} for {Oligomeric} and {Polymeric} {Systems}},
	volume = {22},
	issn = {1549-9618},
	shorttitle = {How {Long} {Is} {Long} {Enough}?},
	url = {https://doi.org/10.1021/acs.jctc.6c00365},
	doi = {10.1021/acs.jctc.6c00365},
	number = {13},
	urldate = {2026-08-10},
	journal = {J. Chem. Theory Comput.},
	author = {Hooven, Natalie E. and Lin, Arthur Y. and Carroll, Charles H. and Cersonsky, Rose K.},
	month = jun,
	year = {2026},
	pages = {6917--6926},
}

@article{hu_efficient_2024,
	title = {Efficient {Machine} {Learning} {Force} {Field} for {Large}-{Scale} {Molecular} {Simulations} of {Organic} {Systems}},
	volume = {7},
	url = {https://www.chinesechemsoc.org/doi/full/10.31635/ccschem.024.202404785},
	doi = {10.31635/ccschem.024.202404785},
	number = {3},
	urldate = {2026-08-10},
	journal = {Chin. Chem. Soc.},
	publisher = {Chinese Chemical Society},
	author = {Hu, Junbao and Zhou, Liyang and Jiang, Jian},
	month = oct,
	year = {2024},
	pages = {716--730},
}

@article{long_polymers_2024,
	title = {Polymers simulation using machine learning interatomic potentials},
	volume = {308},
	issn = {0032-3861},
	url = {https://www.sciencedirect.com/science/article/pii/S0032386124007523},
	doi = {10.1016/j.polymer.2024.127416},
	urldate = {2026-08-10},
	journal = {Polymer},
	author = {Long, Teng and Li, Jia and Wang, Chenlu and Wang, Hua and Cheng, Xiao and Lu, Haifeng and Zhang, Ying and Zhou, Chuanjian},
	month = aug,
	year = {2024},
	pages = {127416},
}

@article{wang_scalable_2021,
	title = {A {Scalable} {Graph} {Neural} {Network} {Method} for {Developing} an {Accurate} {Force} {Field} of {Large} {Flexible} {Organic} {Molecules}},
	volume = {12},
	issn = {1948-7185},
	url = {https://doi.org/10.1021/acs.jpclett.1c02214},
	doi = {10.1021/acs.jpclett.1c02214},
	number = {33},
	urldate = {2026-08-10},
	journal = {J. Phys. Chem. Lett.},
	author = {Wang, Xufei and Xu, Yuanda and Zheng, Han and Yu, Kuang},
	month = aug,
	year = {2021},
	pages = {7982--7987},
}

@book{flory_principles_1953,
	title = {Principles of {Polymer} {Chemistry}},
	url = {http://archive.org/details/dli.ernet.286013},
	urldate = {2026-03-13},
	publisher = {Cornell University Press, New York},
	author = {Flory, Paul J.},
	year = {1953},
}

@article{shinoda_rapid_2004,
	title = {Rapid estimation of elastic constants by molecular dynamics simulation under constant stress},
	volume = {69},
	copyright = {http://link.aps.org/licenses/aps-default-license},
	issn = {1098-0121, 1550-235X},
	url = {https://link.aps.org/doi/10.1103/PhysRevB.69.134103},
	doi = {10.1103/PhysRevB.69.134103},
	language = {en},
	number = {13},
	urldate = {2026-08-13},
	journal = {Phys. Rev. B},
	author = {Shinoda, Wataru and Shiga, Motoyuki and Mikami, Masuhiro},
	month = apr,
	year = {2004},
	pages = {134103},
}

@article{jorgensen_development_1996,
	title = {Development and {Testing} of the {OPLS} {All}-{Atom} {Force} {Field} on {Conformational} {Energetics} and {Properties} of {Organic} {Liquids}},
	volume = {118},
	issn = {0002-7863},
	url = {https://doi.org/10.1021/ja9621760},
	doi = {10.1021/ja9621760},
	number = {45},
	urldate = {2026-08-14},
	journal = {J. Am. Chem. Soc.},
	author = {Jorgensen, William L. and Maxwell, David S. and Tirado-Rives, Julian},
	month = nov,
	year = {1996},
	pages = {11225--11236},
}

@article{siu_optimization_2012,
	title = {Optimization of the {OPLS}-{AA} {Force} {Field} for {Long} {Hydrocarbons}},
	volume = {8},
	issn = {1549-9618},
	url = {https://doi.org/10.1021/ct200908r},
	doi = {10.1021/ct200908r},
	number = {4},
	urldate = {2026-08-14},
	journal = {J. Chem. Theory Comput.},
	author = {Siu, Shirley W. I. and Pluhackova, Kristyna and B{\"o}ckmann, Rainer A.},
	month = mar,
	year = {2012},
	pages = {1459--1470},
}

@article{sherck_end--end_2020,
	title = {End-to-{End} {Distance} {Probability} {Distributions} of {Dilute} {Poly}(ethylene oxide) in {Aqueous} {Solution}},
	volume = {142},
	issn = {0002-7863},
	url = {https://doi.org/10.1021/jacs.0c08709},
	doi = {10.1021/jacs.0c08709},
	number = {46},
	urldate = {2026-08-14},
	journal = {J. Am. Chem. Soc.},
	author = {Sherck, Nicholas and Webber, Thomas and Brown, Dennis Robinson and Keller, Timothy and Barry, Mikayla and DeStefano, Audra and Jiao, Sally and Segalman, Rachel A. and Fredrickson, Glenn H. and Shell, M. Scott and Han, Songi},
	month = nov,
	year = {2020},
	pages = {19631--19641},
}

@article{jeong_mass_2015,
	title = {Mass dependence of the activation enthalpy and entropy of unentangled linear alkane chains},
	volume = {143},
	issn = {0021-9606},
	url = {https://doi.org/10.1063/1.4932601},
	doi = {10.1063/1.4932601},
	number = {14},
	urldate = {2026-08-14},
	journal = {J. Chem. Phys.},
	author = {Jeong, Cheol and Douglas, Jack F.},
	month = oct,
	year = {2015},
	pages = {144905},
}

@book{rubinstein_polymer_2003,
	title = {Polymer {Physics}},
	isbn = {978-0-19-852059-7},
	publisher = {OUP Oxford},
	author = {Rubinstein, Michael and Colby, Ralph H.},
	month = jun,
	year = {2003},
}

@article{mondello_dynamics_1998,
	title = {Dynamics of n-alkanes: {Comparison} to {Rouse} model},
	volume = {109},
	issn = {0021-9606},
	shorttitle = {Dynamics of n-alkanes},
	url = {https://doi.org/10.1063/1.476619},
	doi = {10.1063/1.476619},
	number = {2},
	urldate = {2026-08-14},
	journal = {J. Chem. Phys.},
	author = {Mondello, Maurizio and Grest, Gary S. and Webb, III, Edmund B. and Peczak, P.},
	month = jul,
	year = {1998},
	pages = {798--805},
}

@article{li_critical_1995,
	title = {Critical exponents, hyperscaling, and universal amplitude ratios for two- and three-dimensional self-avoiding walks},
	volume = {80},
	issn = {1572-9613},
	url = {https://doi.org/10.1007/BF02178552},
	doi = {10.1007/BF02178552},
	language = {en},
	number = {3},
	urldate = {2026-08-14},
	journal = {J. Stat. Phys.},
	author = {Li, Bin and Madras, Neal and Sokal, Alan D.},
	month = aug,
	year = {1995},
	pages = {661--754},
}

\end{document}